\documentclass{elsart}

\usepackage{amsmath}
\usepackage{amssymb}
\usepackage{latexsym}
\usepackage{graphicx}

\journal{Astroparticle Physics}

\begin{document}

\begin{frontmatter}

\title{Probing Quantum Decoherence in Atmospheric Neutrino Oscillations
with a Neutrino Telescope}

\author[Applied]{Dean Morgan}
\author[Applied]{Elizabeth Winstanley\corauthref{cor}}
\ead{E.Winstanley@sheffield.ac.uk}
\author[Marseille]{Jurgen Brunner}
\author[Physics]{Lee F. Thompson}

\address[Applied]{Department of Applied Mathematics,
The University of Sheffield, Hicks Building, Hounsfield Road,
Sheffield, S3 7RH, U.K.}
\address[Marseille]{Centre de Physique des Particules de Marseille,
163 Avenue de Luminy - \\ Case 907, 13288 Marseille Cedex 09, France.}
\address[Physics]{Department of Physics and Astronomy,
The University of Sheffield,\\ Hicks Building, Hounsfield Road, Sheffield, S3 7RH, U.K.}

\corauth[cor]{Corresponding author}

\begin{abstract}
Quantum decoherence, the evolution of pure states into mixed
states, may be a feature of quantum gravity.
In this paper, we show how these effects can be modelled for atmospheric neutrinos and
illustrate how the standard oscillation picture is modified.
We examine how neutrino telescopes, such as ANTARES, are
able to place upper bounds on these quantum decoherence effects.
\end{abstract}


\end{frontmatter}

\section{Introduction}
\label{sec:intro}

Quantum field theory and general relativity are supremely successful theories, both having been
experimentally verified to great accuracy.
However, these two theories are fundamentally incompatible:
general relativity is non-renormalizable
as a quantum field theory.
Many candidate theories of quantum gravity exist in the literature, but there is as yet no
consensus as to which will provide the final fundamental theory.
For a recent review of the current state of play, see \cite{smolin}.
Quantum gravity becomes particularly important in two realms of physics: the very early universe
(prior to the Planck time, $10^{-43}$ s); and in black holes, when gravitational fields are
strong but quantum effects cannot be ignored.

There are two different approaches to uncovering the theory of
quantum gravity: firstly, there is the `top down' method, which involves
writing a theory down and then attempting to see if the theory can
make any physical predictions which are testable.
The second, which is the method embraced here, is the `bottom up' method, in
which we analyze experimental data and attempt to fit the data to
various phenomenological models (for a recent review, see, for example, \cite{amelino1}).
This way, we may narrow down the number of models
until, eventually, a single theory emerges, which is consistent
with experiment.
There are various ways in which quantum gravity may alter fundamental physics;
the final theory will no doubt include highly unexpected effects on
basic physical principles.
Here we are concerned only with one such mechanism: the influence of
quantum gravity on standard quantum mechanical time-evolution.

Naively, one might expect that quantum gravity is beyond the reach
of experimental physics, since its relevant energy scale is the
Planck energy, approximately $10^{19}$ GeV.
However, recent
theoretical models coming from string theory suggest that the
energy scale of quantum gravity may be as low as a few TeV \cite{rs}.
Moreover, even with more conservative estimates of the
quantum gravity energy scale, comparatively low energy physics may
be sensitive to quantum gravity effects, for example, through
quantum-gravity induced modifications of quantum mechanics
resulting in quantum decoherence \cite{EHNS}.

In this article we are concerned with one such low-energy system:
atmospheric neutrinos.
This is a comparatively simple quantum
system, which lends itself to an analysis of quantum decoherence
effects.
There is already a body of work on the theoretical
modelling of quantum decoherence effects on neutrino oscillations
\cite{benatti,liu,ma,chang,klapdor}.
However, the only analysis of decoherence effects in
experimental data relating to $\nu _{\mu } \rightarrow \nu _{\tau }$
oscillations available in the literature to date is for
Super-Kamiokande and K2K data \cite{lisi,fogli,SKspectra}.
An initial analysis \cite{lisi} was able to put bounds on quantum decoherence
effects in a simple model.
Subsequently, a detailed analysis of the same simple model
\cite{fogli} found that quantum decoherence effects were slightly
disfavoured compared with the standard oscillation scenario, but
that they could not be completely ruled out.
Further experimental work is therefore needed, and here we shall study the sensitivity
of the ANTARES neutrino telescope \cite{ANTARES} to these quantum
decoherence effects in atmospheric neutrino oscillations.
Although neutrino telescopes such as ANTARES are primarily designed for the
detection and study of neutrinos of astrophysical and cosmological origin,
atmospheric neutrinos form the most important background to such sources and
are likely to provide the first physics results.

As well as the simple model considered in Refs. \cite{lisi,fogli},
in this paper we shall study models which
are more exotic than those considered previously, as they include
effects such as non-conservation of energy (within the neutrino
system).
Furthermore, while these models have been studied
theoretically in the literature
\cite{benatti,liu,ma,chang,klapdor}, they have not been
experimentally tested.
It should be
emphasized that such quantum decoherence is not the only possible
effect of quantum gravity on neutrinos \cite{nick}.
For example, quantum
gravity is expected to change the energy dependence of the
oscillation length \cite{brustein}, and also alter the dispersion
relation \cite{hugo}, leading to observable effects \cite{king}.
We shall also briefly discuss other, non-quantum gravity, possible
corrections to atmospheric neutrino oscillations and how they
relate to our models here, as well as comparing our results on
quantum decoherence parameters with those coming from other
experimental (i.e. non-neutrino) systems (see, for example,
\cite{ELMN,benattibound}).
Finally, although our simulations are
specific to ANTARES, we anticipate that other high energy neutrino
experiments such as AMANDA \cite{AMANDA}, IceCube \cite{ICECUBE}
and NESTOR \cite{NESTOR}, may well be able to probe these effects.

Given the high energy and long path-length of atmospheric neutrinos studied
by ANTARES, this study is complementary to ``long-baseline'' neutrino
oscillation experiments.
As well as the analysis of K2K data \cite{fogli},
a related analysis of KAMLAND data has also been performed in \cite{KAMLANDspectra}.
In both cases decoherence is disfavoured over the standard oscillation picture.
The potential for bounding quantum decoherence and other damping signatures in future
reactor experiments, such as MINOS \cite{MINOS} and OPERA \cite{OPERA},
is discussed in \cite{blennow}.

The outline of this paper is as follows.
In section \ref{sec:qg}, we discuss how quantum gravity may be expected
to modify quantum mechanics, before applying the formalism developed
to atmospheric neutrino oscillations in section \ref{sec:atmqg}.
We study in detail some specific models of quantum decoherence in section \ref{sec:sensitivity},
and describe the results of ANTARES sensitivity simulations for these models
in section \ref{sec:oscfit}.
The projected ANTARES sensitivities are compared with results from other experiments
(both neutrino and non-neutrino) in section \ref{sec:exp}, and we briefly
discuss other possible corrections to atmospheric neutrino oscillations in
section \ref{sec:other}.
Finally, our conclusions are presented in section \ref{sec:conc}.
Unless otherwise stated, we use units in which $c=\hbar =1$.

\section{Quantum gravity induced modifications of quantum mechanics}
\label{sec:qg}

In quantum gravity, space-time is expected to adopt a `foamy' structure \cite{wheeler}:
tiny black holes form out of the vacuum (in the same
way that quantum particles can be formed out of the vacuum).
These will be very short-lived, evaporating quickly by Hawking radiation \cite{hawking1}.
Each microscopic black hole may induce variations in the standard time-evolution
of quantum mechanics, as an initially pure quantum state (describing the space-time
before the formation of the black hole) can evolve to a mixed quantum state
(describing the Hawking radiation left after the black hole has evaporated).
In particular, in quantum gravity it is no longer necessarily the case
that pure states cannot evolve to mixed states.

In order to describe this process, Hawking suggested a quantum
gravitational modification of quantum mechanics \cite{hawking2},
which is based on the density matrix formalism.
An initial quantum state is described by a density
matrix $\rho _{in}$, and this state then evolves to a final
state described by a density matrix $\rho _{out}$.
The two density matrices are related by
\begin{displaymath}
\rho_{out} = \$\rho_{in},
\end{displaymath}
where $ \$ $ is an operator known as the {\em {super-scattering}} operator.
In ordinary quantum mechanics, the super-scattering operator
can be factorized as
\begin{equation}
\$ = \mathcal{S}\mathcal{S}^{\dag},
\label{scat:fac}
\end{equation}
where ${\mathcal {S}}$ is the usual $S$-matrix.
In this case, a pure initial state always leads to a pure final
state.
However, in quantum gravity it is no longer necessarily the
case that the super-scattering matrix can be
factorized as in equation (\ref{scat:fac}),
and, if it cannot, then the evolution
from pure state to mixed state is allowed.
The disadvantage of this approach is that it requires the construction of
asymptotic ``in'' and ``out'' states.

Therefore, here we will take an alternative approach, which is more
useful for our application to atmospheric neutrino oscillations,
allowing the evolution of pure to mixed states by modifying the
differential equation describing the time-evolution of the density matrix $\rho $ \cite{EHNS}.
In standard quantum mechanics this differential equation takes the form
\begin{displaymath}
\dot{\rho} = -i[H,\rho],
\end{displaymath}
and could be modified by quantum gravity to the equation \cite{EHNS}
\begin{equation}
\dot{\rho} = -i[H,\rho]+\delta\!\!\,\!\!\slash
H\rho.
\label{dens:matnp}
\end{equation}
In the above equations, $H$ is the Hamiltonian of the system,
$\delta\!\!\,\!\!\slash H$ is the most general linear extension
which maps hermitian matrices to hermitian matrices and the dot
denotes differentiation with respect to time.
Various conditions need to be imposed on $\delta\!\!\,\!\!\slash H$
to ensure that probability is still conserved and that ${\rm {Tr}} \rho ^{2}$ is
never greater than unity \cite{EHNS}.
These conditions will be implemented in the next section when we apply this
approach to atmospheric neutrinos.
However, the precise form of $\delta\!\!\,\!\!\slash H$ would need to come
from a complete, final, theory of quantum gravity.
In this paper we take a phenomenological approach, modelling $\delta\!\!\,\!\!\slash H$
in a manner independent of its origin in quantum gravity.
Ultimately one may hope that experimental observations may be able to
constrain the form of $\delta\!\!\,\!\!\slash H$ and hence its theoretical origin.

One form of $\delta\!\!\,\!\!\slash H$ which is frequently employed in the literature is the
Lindblad form~\cite{lindblad}
\begin{equation}
\delta\!\!\,\!\!\slash H \rho = \sum _{n} \left[
D_{n} , \left[ D_{n}, \rho \right] \right] ,
\label{eq:lindblad}
\end{equation}
where the $D_{n}$ are self-adjoint operators which commute with the Hamiltonian of the theory.
Such modifications of standard quantum-mechanical time evolution have been widely studied,
both as a solution to the quantum measurement problem \cite{GRW} and for decoherence due
to the interaction with an environment \cite{decobook}.
In recent work in discrete quantum gravity, it has been shown that
$\delta\!\!\,\!\!\slash H$ has the Lindblad form with a single operator $D$ which is
proportional to the Hamiltonian $H$ \cite{gambini}.
However, it should be emphasized that this is not the only possible approach to
quantum gravity, and that other formulations indicate that in general
$\delta\!\!\:\!\!\slash H$ does not have the Lindblad form.
This is the reason for our more general set-up in this paper.

Modifications of quantum mechanics (whether from quantum gravity
or other origins) of the form (\ref{dens:matnp}) are often known
as {\em {quantum decoherence}} effects as they emerge from
dissipative interactions with an environment (of which space-time
foam is just one example), and allow transitions from pure to
mixed quantum states (i.e. a loss of {\em {coherence}}). Such
effects have been studied in quantum systems other than
atmospheric neutrinos, for example, the neutral kaon system, for
which there exists an extensive literature (see, for example,
\cite{ELMN,benattibound}).

It should be stressed that although our primary motivation for studying
quantum decoherence is from quantum gravity, in fact our modelling in the
subsequent sections is independent of the source of the quantum decoherence.
Furthermore, Ohlsson \cite{ohlsson} has discussed how Gaussian uncertainties in
the neutrino energy and path length, when averaged over, can lead to
modifications of the standard neutrino oscillation probability which are similar
in form to those coming from quantum decoherence, and which are therefore included in our
analysis.
See also \cite{barenboim1} for further discussion of these effects.

\section{Quantum decoherence modifications of atmospheric neutrino oscillations}
\label{sec:atmqg}

We now consider how quantum gravity induced modifications of quantum mechanics,
as outlined in the previous section, may affect atmospheric neutrino oscillations.
The quantum system of interest is then simply made up of muon and tau neutrinos.
The mathematical modelling of quantum decoherence effects in this system is not dissimilar
to that for the neutral kaon system \cite{EHNS,ELMN}.
In this paper we will assume that quantum decoherece affects neutrinos and anti-neutrinos
in the same way, and do not consider the possibility that quantum decoherence may appear
only in the anti-neutrino sector \cite{barenboim2}.

In order to implement equation (\ref{dens:matnp}) for the system
of atmospheric muon and tau neutrinos, we need to
represent the matrices $H$ and $\rho$ in terms of a specific basis
which comprises the standard Pauli matrices $\sigma _{i}$.
We may, therefore, write the density matrix, Hamiltonian and additional term
$\delta\!\!\:\!\!\slash H$ in
terms of the Pauli matrices as
\begin{displaymath}
  \rho = \frac{1}{2}\rho_{\mu }\sigma_{\mu };
  \quad\quad
  H = \frac{1}{2}h_{\nu }\sigma_{\nu };
  \quad \quad
  \delta\!\!\:\!\!\slash H = \frac {1}{2} h'_{\kappa }\sigma _{\kappa };
\end{displaymath}
where the Greek indices run from 0 to 3, and a summation over repeated indices is understood.
Similarly decomposing the time derivative of the density matrix,
\begin{displaymath}
{\dot {\rho }} = \frac {1}{2} {\dot {\rho }}_{\mu } \sigma _{\mu } ,
\end{displaymath}
where, using equation (\ref{dens:matnp}):
\begin{equation}
\dot{\rho}_{\mu } = (h_{\mu \nu } +
h'_{\mu \nu })\rho_{\nu }.
\label{eq:dotrho}
\end{equation}
Here, $h$ represents the standard atmospheric neutrino oscillations and
$h'$ the quantum decoherence effects.
In order to preserve unitarity, the first row and column of the matrix $h'$ must vanish;
the fact that ${\rm {Tr}} \rho ^{2}$ can never exceed unity means that $h'$ must be
negative semi-definite \cite{EHNS}.
The most general form of $h'$ is therefore given by
\begin{equation}
h' = -2\left(%
\begin{array}{cccc}
  0 & 0 & 0 & 0 \\
  0 &  a & b & d \\
  0 & b & \alpha  & \beta \\
  0 & d & \beta & \delta \\
\end{array}%
\right) ;
\label{eq:hprime}
\end{equation}
where $a$, $b$, $d$, $\alpha$, $\beta$ and $\delta$ are real
constants parameterizing the quantum decoherence effects.

Particularly with regard to quantum decoherence induced by
space-time foam (see section \ref{sec:qg}), it is not clear
exactly what assumptions about the time-evolution matrix $h'$ are
reasonable, for example, it may no longer be the case that energy
is conserved within the neutrino system, as energy may be lost to
the environment of space-time foam.
The first observation is that
$h'$ has non-zero entries in the final row and column ($d$, $\beta
$ and $\delta $), which correspond to terms which violate energy
conservation \cite{EHNS}.
The matrix $h'$ can be further
simplified by making additional assumptions.
For example, since
the eigenvalues of the density matrix $\rho $ correspond to
probabilities, we require that these are positive (this is known
as {\em {simple positivity}}).
However, we can make a much
stronger assumption known as {\em {complete positivity}},
discussed in \cite{benatti}.
It arises in the quantum mechanics of
open systems, and ensures the positivity of the density matrix
describing a much larger system, in which the neutrinos are
coupled to an external system.
Complete positivity is a strong
assumption, favoured by mathematicians because it is a powerful
tool in proving theorems related to quantum decoherence.
For the matrix $h'$, assuming complete
positivity leads to a number of inequalities on the quantum
decoherence parameters \cite{benatti}, and assuming both complete
positivity and energy conservation within the neutrino system
means that $a=\alpha $ and all other quantum decoherence
parameters must vanish \cite{benatti} (which gives the
Lindblad form (\ref{eq:lindblad})).
It is this simplified model
which has been most studied in the literature to date
\cite{lisi,fogli}.
However, in this paper we wish to consider not
only this simplified model but other more general models of
quantum decoherence which do not necessarily satisfy complete
positivity or energy conservation.

Substituting (\ref{eq:hprime}) into (\ref{eq:dotrho}),
and incorporating the standard atmospheric neutrino oscillations
in the matrix $h$ (\ref{eq:dotrho}), we obtain
the following differential equations for the time-evolution of the
components of the density matrix:
\begin{eqnarray}
\dot{\rho}_{0} &=& 0;
\nonumber \\
\dot{\rho}_{1} &=& -2a \rho_{1} - 2\left( b- \frac{\Delta
m^{2}}{4E} \right) \rho_{2} -2 d \rho _{3};
\nonumber \\
\dot{\rho}_{2} &=& -2 \left( b+ \frac{\Delta m^{2}}{4E}
\right)\rho_{1} -2\alpha \rho_{2} -2 \beta \rho _{3};
\nonumber \\
\dot{\rho}_{3} &=&
-2d \rho _{1} -2\beta \rho _{2} -2 \delta \rho _{3};
\label{diffeqn:rho}
\end{eqnarray}
where $\Delta m^{2}$ is the usual difference in squared masses of the neutrino mass eigenstates,
and $E$ is the neutrino energy.

By integrating the differential equations (\ref{diffeqn:rho}) with suitable initial
conditions, the probability for a muon neutrino to oscillate into a tau neutrino can
be calculated.
In general, this
probability has the form~\cite{benatti}:
\begin{eqnarray}
P\left[ \nu _{\mu } \rightarrow \nu _{\tau } \right]
 &  = & \frac {1}{2} \left\{1-\cos^{2}(2\theta)
 M_{33}(E,L)-\sin^{2}(2\theta) M_{11}(E,L)
\right.
\nonumber
\\ & & \left.
- \frac {1}{2} \sin 4\theta \left[
M_{13}(E,L) + M_{31}(E,L) \right] \right\} ,
\label{eq:genprob}
\end{eqnarray}
where the functions $M_{11}(E,L)$, $M_{33}(E,L)$, $M_{13}(E,L)$ and $M_{31}(E,L)$ are
elements of the matrix $M(E,L)$ which is defined as
\begin{equation*}
M(E,L) = \exp \left[-2{\mathcal {H}}(E)L\right] ,
\end{equation*}
where
\begin{equation*}
{\mathcal {H}} (E) = \left(
\begin{array}{ccc}
a & b-\frac {\Delta m^{2}}{4E} & d \\
b+ \frac {\Delta m^{2}}{4E} & \alpha & \beta  \\
d & \beta & \delta
\end{array}
\right) .
\end{equation*}
In general there is no simple closed form expression for the elements of the matrix
$M(E,L)$.
Here we shall study various special cases in which there will be closed form
functions of the new parameters appearing in the probability (\ref{eq:genprob}).
For each model, we consider
three possible ways in which the quantum decoherence parameters can depend on the
energy $E$ of the neutrinos  (we
shall assume that where there are two or more quantum decoherence parameters, they have the
same energy dependence):
\begin{enumerate}
\item
Firstly, the simplest model is to suppose that the quantum decoherence parameters are constants,
and have no energy dependence.
In this case we write
\begin{displaymath}
\alpha = \frac {1}{2} \gamma _{\alpha } ,
\end{displaymath}
with similar expressions for the other quantum decoherence
parameters.
\item
Secondly, we consider the situation in which the
quantum decoherence parameters are inversely proportional to the
energy $E$, which leads to probabilities which are Lorentz
invariant. We define the constants of proportionality $\mu _{\alpha }$ etc. according
to the equation:
\begin{displaymath}
\alpha = \frac {\mu ^{2}_{\alpha }}{4E} .
\end{displaymath}
This energy dependence has received the most attention to date in
the literature \cite{lisi,fogli}.
We should comment that Lorentz invariance is not necessarily expected to hold exactly in quantum
gravity \cite{hugo}.
\item
Finally, we study a model
suggested in reference \cite{EHNS}, and
which arises in some semi-classical calculations of decoherence phenomena in black hole
and recoiling D-brane geometries \cite{winstanley}, namely that the
quantum decoherence parameters are proportional to the energy
squared, so that the quantum decoherence parameters are given by
expressions of the form
\begin{displaymath}
\alpha = \frac {1}{2} \kappa _{\alpha } E^{2} ,
\end{displaymath}
where $\kappa _{\alpha }$ is a constant.
This energy dependence also arises from discrete quantum gravity \cite{gambini}, where
the operator $D$ in the Lindblad form (\ref{eq:lindblad}) is proportional to the Hamiltonian.
On dimensional grounds, the constant of proportionality $\kappa $ contains a factor of $1/M_{P}$,
where $M_{P}$ is the Planck mass, $10^{28}$ eV.
This suppression of quantum gravity effects by a single power of the Planck mass is
naively unexpected (as it corresponds to a dimension-5, non-renormalizable operator in the theory,
see, for example \cite{myers}).
One would instead expect quantum gravity effects to be suppressed by the Planck mass squared.
\end{enumerate}
We should also point out that an alternative energy dependence, namely that
quantum decoherence parameters are proportional to
\begin{equation*}
\frac {\left( \Delta m^{2} \right) ^{2}}{E^{2}}
\end{equation*}
has been suggested from $D$-brane interactions in non-critical
string theory \cite{nick,adler}.
However, we have not been able to
derive any meaningful results for this energy dependence, in
accordance with the observation in \cite{nick} that it is unlikely
that current, or even planned, experiments will be able to probe
this energy dependence.

\begin{figure}
\begin{center}
\includegraphics[angle=270,width=8cm]{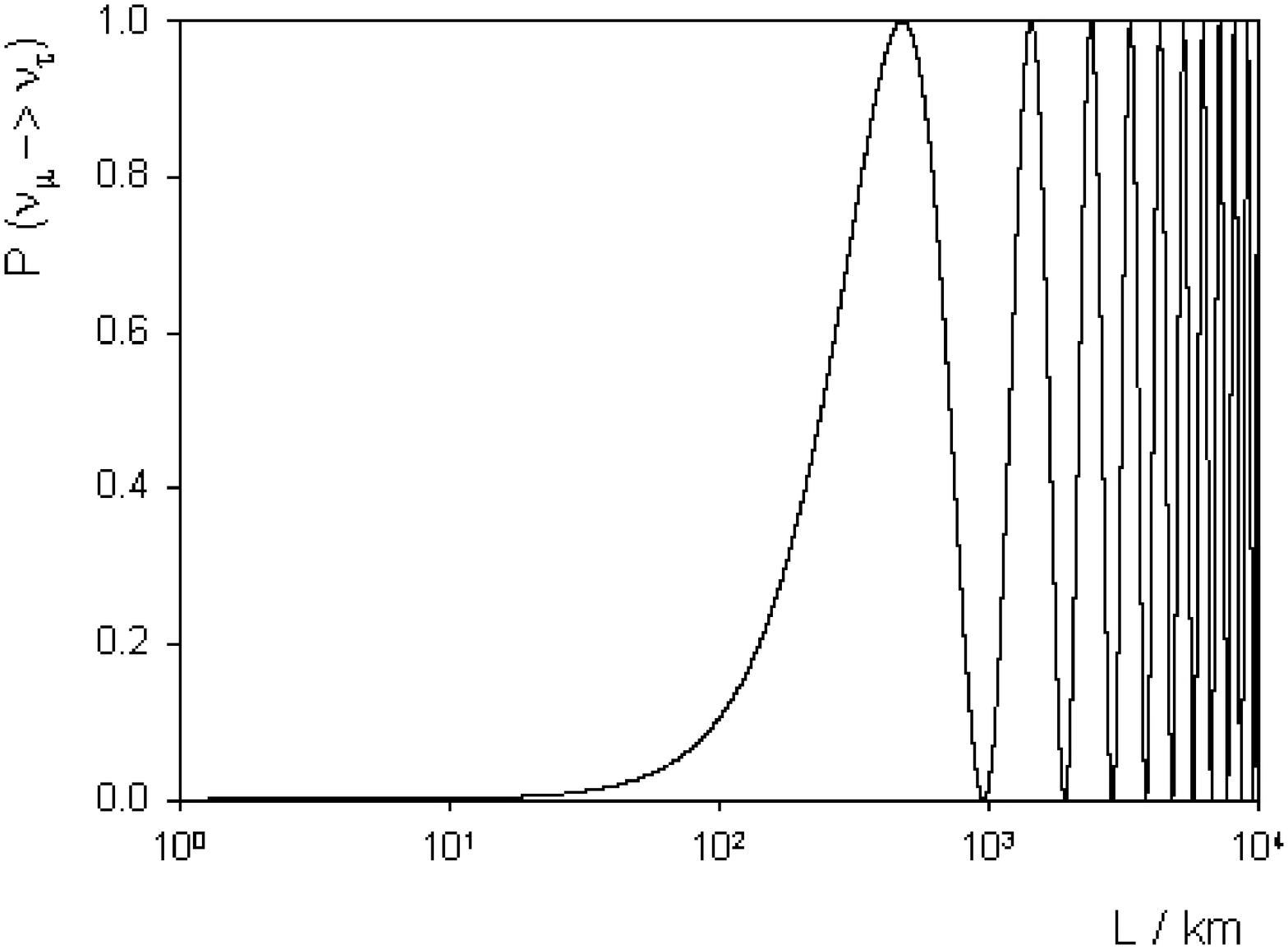}\\
\includegraphics[angle=270,width=8cm]{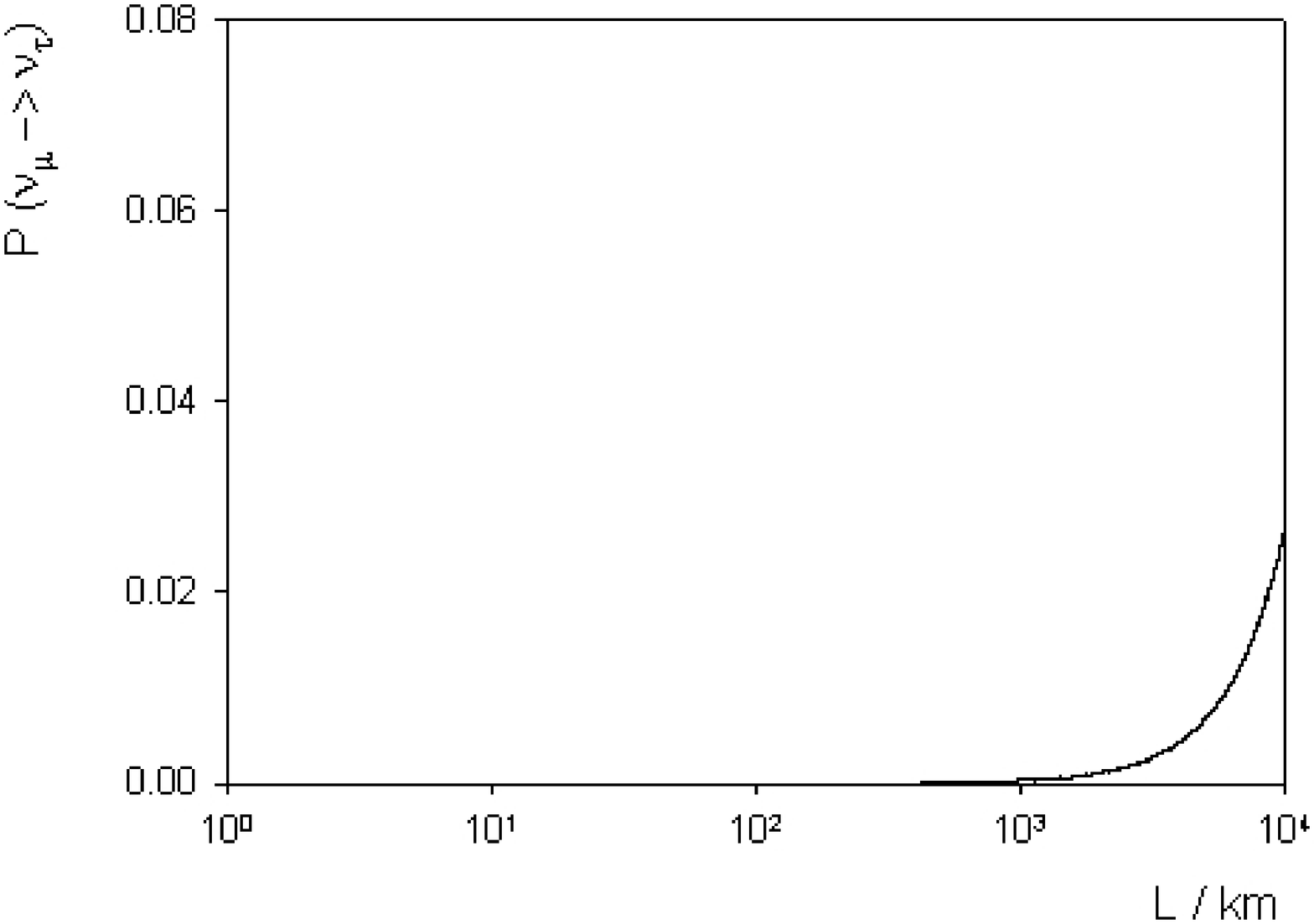}
\caption{Standard atmospheric neutrino oscillation probability
(\ref{eq:standard}) as a function of path length $L$ (measured in
km), for fixed neutrino energy $E$ equal to 1 GeV (top) and 200
GeV (bottom).}
\label{fig:standard1}
\end{center}
\end{figure}

\begin{figure}
\begin{center}
\includegraphics[angle=270,width=8cm]{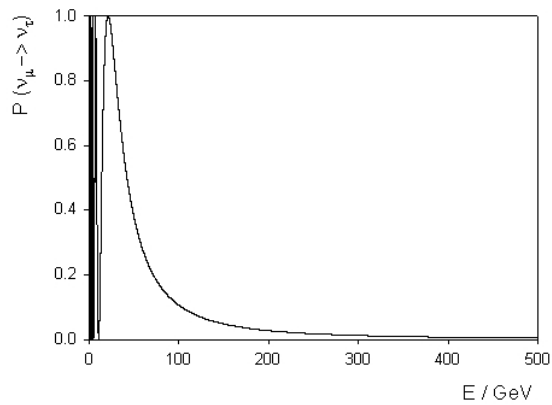}\\
\caption{Standard atmospheric neutrino oscillation probability
(\ref{eq:standard}) as a function of neutrino energy $E$ (measured
in GeV), for fixed path length equal to  10000 km.}
\label{fig:standard2}
\end{center}
\end{figure}

To see how quantum decoherence affects atmospheric neutrino oscillations, we
now compare the oscillation probabilities.
Although the oscillation probabilities are unobservable directly, they
nonetheless provide insight into the underlying physics behind
observable neutrino behaviour.
Firstly, the standard atmospheric neutrino oscillation probability is given by
\begin{equation}
P[\nu _{\mu } \rightarrow \nu _{\tau} ] = \frac{1}{2} \left[
1-\cos \left( 6.604\times 10^{-3}\frac {L}{E}\right) \right] ;
\label{eq:standard}
\end{equation}
where we have restored the constants $c$ and $\hbar$ and the
neutrino energy $E$ is measured in GeV with the path length $L$
being measured in km.
In equation (\ref{eq:standard}), we have
taken $\Delta m^{2}$ and $\sin ^{2}(2\theta )$ to have their
best-fit values \cite{fogli}
\begin{equation}
\Delta m^{2} = 2.6\times 10^{-3} {\mbox {eV${}^{2}$}} ;
\qquad
\sin ^{2} (2\theta ) = 1.
\label{eq:bestfit}
\end{equation}
The oscillation probability
(\ref{eq:standard}) is plotted in figure \ref{fig:standard1}
as a function of path length $L$, for fixed neutrino energy $E$ equal
to 1 GeV and 200 GeV, and in figure \ref{fig:standard2} as a function of neutrino energy
$E$ for fixed path length $L$ equal to 10000 km.
Although atmospheric neutrino experiments to date have tended to focus on
neutrino energies of the order of a few GeV \cite{lisi,fogli},
ANTARES is sensitive to high energy neutrinos (above the order of tens of GeV).
Hence we are particularly interested in the plots for energies of 200 GeV although we have
included 1 GeV for comparison purposes.
For further discussion of the atmospheric neutrino flux at high energies, see,
for example, references \cite{atmflux}.
Since ANTARES detects neutrinos which have passed through the Earth, we have chosen our reference
path length to be a similar order of magnitude to the radius of the Earth.

We now plot the corresponding probabilities including quantum
decoherence effects, the general oscillation probability being
given by equation (\ref{eq:genprob}).
For illustration purposes,
we take values of the quantum decoherence parameters equal to the
upper bounds of the sensitivity regions found in table
\ref{table:gammukap} in section \ref{sec:exp}, except that we take
$d=\beta=\delta =0$ as these parameters have a negligible effect
on the probability.

\begin{figure}
\begin{center}
\includegraphics[angle=270,width=8cm]{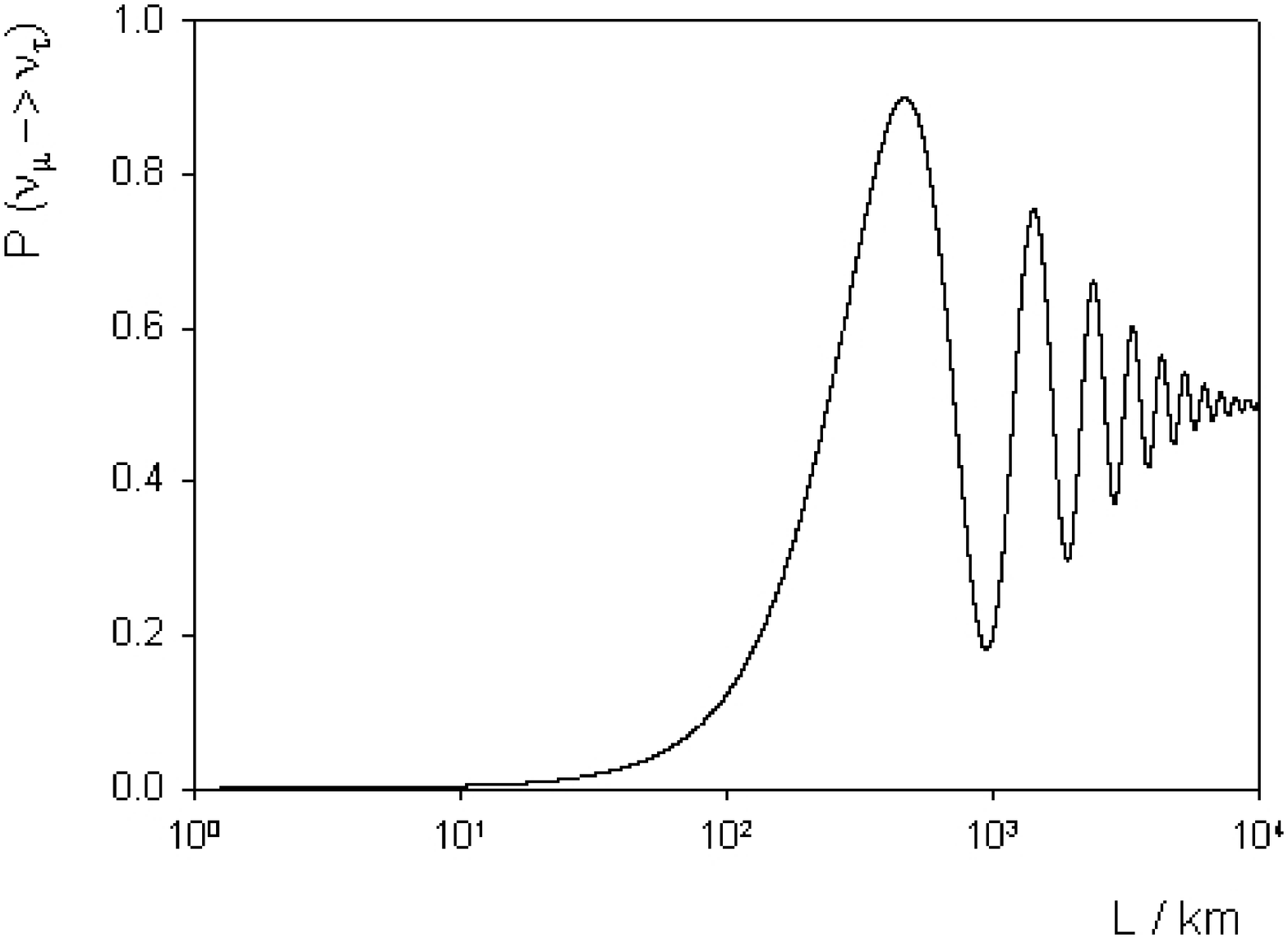}\\
\includegraphics[angle=270,width=8cm]{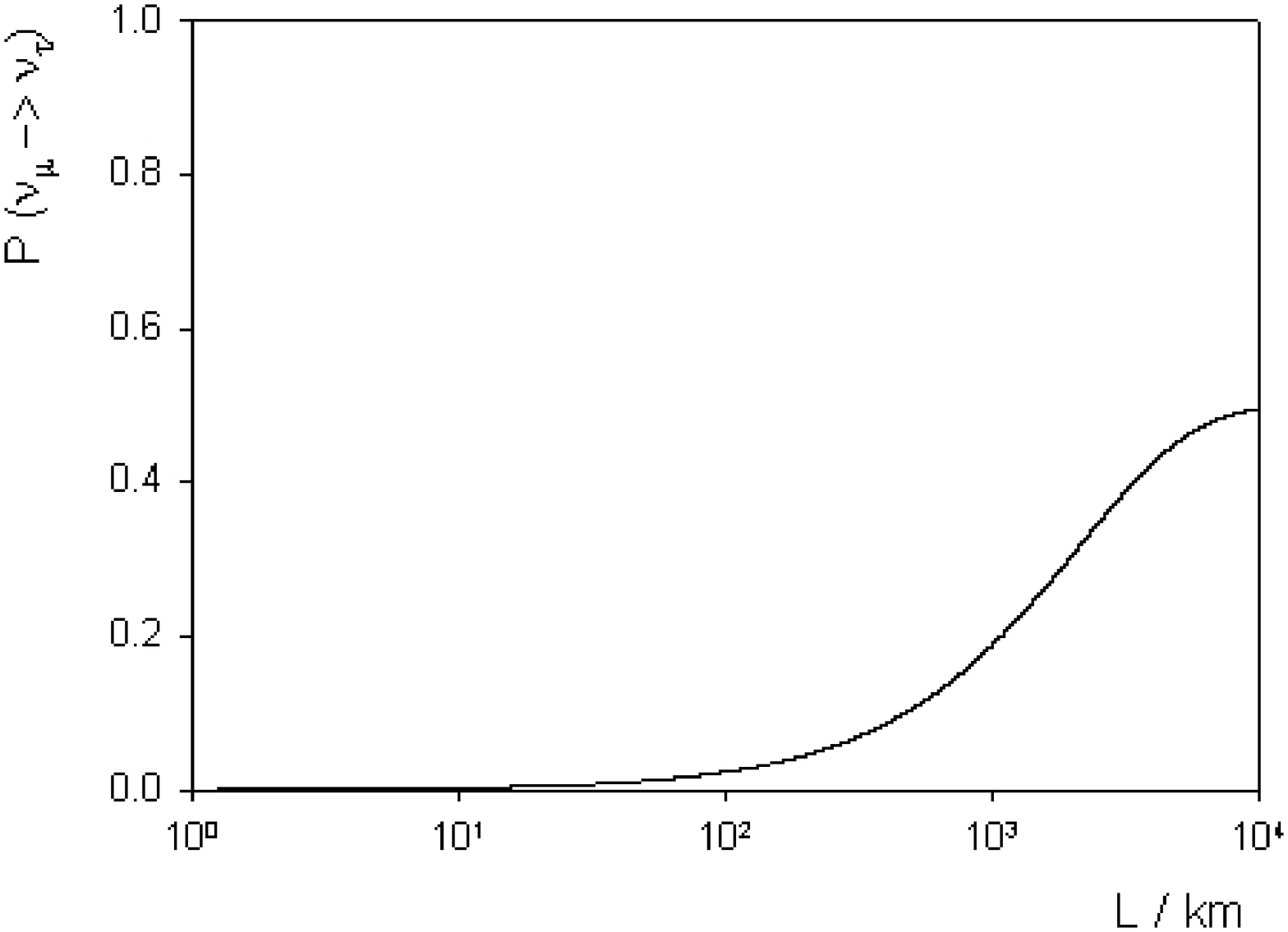}
\caption{Atmospheric neutrino oscillation probability (\ref{eq:probgamma}),
including quantum decoherence parameters which are independent of the neutrino energy,
plotted as a
function of path length $L$ (measured in km), for fixed neutrino energy
$E$ equal to 1 GeV (top) and 200 GeV (bottom).}
\label{fig:gamma1}
\end{center}
\end{figure}

\begin{figure}
\begin{center}
\includegraphics[angle=270,width=8cm]{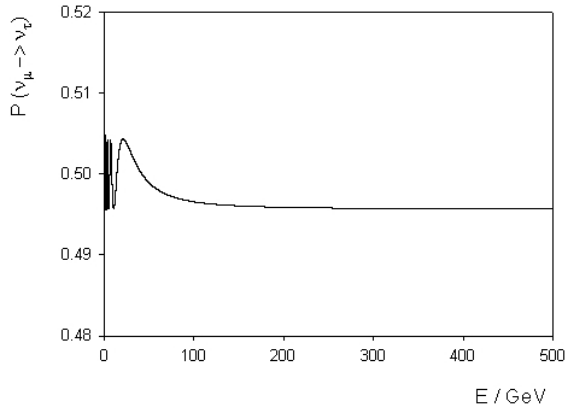}\\
\caption{Atmospheric neutrino oscillation probability
(\ref{eq:probgamma}), including quantum decoherence parameters which are
independent of the neutrino energy, plotted
as a function of neutrino energy $E$ (measured
in GeV), for fixed path length equal to  10000 km.}
\label{fig:gamma2}
\end{center}
\end{figure}

Consider firstly the case in which the quantum decoherence
parameters do not depend on the neutrino energy, and with the
values of the decoherence parameters taken from table
\ref{table:gammukap}, the oscillation probability is:
\begin{eqnarray}
P \left[ \nu _{\mu } \rightarrow \nu _{\tau } \right] &  = &
\nonumber \\
& & \hspace{-2cm}
\frac{1}{2}\left\{ 1 - e^{-4.75\times10^{-4}L} \cos\left[
2L\left(\frac{1.09\times10^{-5}}{E^{2}}-2.723\times10^{-12}\right) ^{\frac {1}{2}}
\right] \right\}.
\label{eq:probgamma}
\end{eqnarray}
This probability (\ref{eq:probgamma}) is plotted in figure \ref{fig:gamma1} as a function
of path length $L$ for two fixed values of neutrino energy $E$ (the same as those in figure
\ref{fig:standard1}), and in figure \ref{fig:gamma2} as a function of energy $E$ for a
fixed value of path length $L$ (again, the same as in figure \ref{fig:standard2}).
The graphs in figures \ref{fig:gamma1} and \ref{fig:gamma2} should be compared with
those in figures \ref{fig:standard1} and \ref{fig:standard2} respectively.
The effect of the quantum decoherence can be clearly seen.
At low energy ($E=1$ GeV), the standard oscillations are damped by quantum decoherence
until the oscillation probability converges to one half (corresponding
to complete decoherence) at a path length of about
$10^{4}$ km.
However, at high energy ($E=200$ GeV), the difference caused by quantum
decoherence is highly marked (note the different vertical scales in the
graphs in figures \ref{fig:standard1} and \ref{fig:gamma1}), with a large enhancement
of the oscillation probability.
This enhancement of the oscillation probability at large path lengths can also be
seen by comparing figures \ref{fig:standard2} and \ref{fig:gamma2}.
For standard oscillations, the probability tends to zero for fixed path length
and increasing energy, but this is not the case when quantum decoherence is included,
the damping factor in the oscillation probability (\ref{eq:probgamma}) ensuring
that the oscillation probability tends to a limiting value for large energy at fixed path-length.

\begin{figure}
\begin{center}
\includegraphics[angle=270,width=8cm]{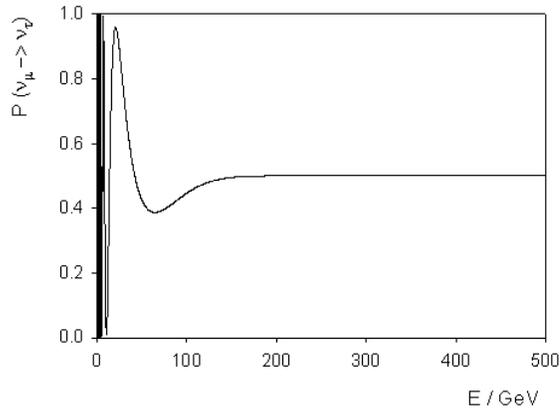}\\
\caption{Atmospheric neutrino oscillation probability
including quantum decoherence parameters which are
proportional to the neutrino energy squared, plotted
as a function of neutrino energy $E$ (measured
in GeV), for fixed path length equal to 10000~km.}
\label{fig:kappa2}
\end{center}
\end{figure}

For larger values of the decoherence parameters, these effects become even more marked,
and, for sufficiently large decoherence parameters, the oscillations may be damped away
completely, even at low energy.
This is in effect what has happened at high energy in figure \ref{fig:gamma1}.
As would be expected, as we decrease the magnitudes of the decoherence parameters,
the decoherence effects become less significant and the probabilities become
indistinguishable from those for standard neutrino oscillations.
However, even for very small quantum decoherence parameters, there may be significant
effects at high energy.
For example, in figure \ref{fig:kappa2}, we plot the oscillation probability
for fixed path length and varying energy
for the case in which the quantum decoherence parameters are proportional
to the neutrino energy squared.
In this situation, the bounds we find in section \ref{sec:oscfit} are very strong,
and the oscillation probability for low energy neutrinos is indistinguishable
from that for standard neutrino oscillations.
However, as can be seen in figure \ref{fig:kappa2}, there is a significant
change in the neutrino oscillation probability at high energy.

The values of the quantum decoherence parameters used in producing figure \ref{fig:kappa2}
are sufficiently small that they have not been ruled out by current atmospheric neutrino
oscillation data \cite{lisi}.
It is clear from figure \ref{fig:kappa2} that high energy neutrinos are necessary in order
to probe such quantum decoherence effects, and it is the main result of this paper
that neutrino telescopes such as ANTARES are indeed able to perform this, in contrast
to comparatively low-energy experiments.
We would expect the bounds we discuss in section \ref{sec:oscfit}, attainable with
atmospheric neutrino data, to be further tightened with neutrinos of astrophysical
rather than atmospheric origin \cite{dan}.

\section{Specific Models of Quantum Decoherence}
\label{sec:sensitivity}

In this section we now discuss in detail specific models of
quantum decoherence.
The most general time evolution equation
including quantum decoherence effects (\ref{eq:hprime}) involves six parameters
additional to those for the standard neutrino
oscillations. It is not possible to derive analytically
in a simple closed form a general
oscillation probability involving all six extra parameters, and either a
numerical approach or an approximation scheme would be necessary \cite{benatti,liu}.
Furthermore, it is unlikely that, at least to begin with,
experimental data will be able to produce meaningful bounds/fits
for all six additional parameters.
In view of the above, we shall
consider models in which only two or fewer of the quantum
decoherence parameters $\alpha $, $\beta $, $\delta $, $a$, $b$,
or $d$ are non-zero, but we study different combinations of these
parameters in order to incorporate different possible effects.

Various models have been considered in the literature to date
(see, for example, \cite{benatti,liu,ma,chang,klapdor}), although
the only work of which we are aware involving analysis of
experimental data is \cite{lisi,fogli}.
The models which have
received most attention to date are those with $a$, $\alpha $ and
$b$ non-zero but $d$, $\beta $ and $\delta $ zero
\cite{benatti,ma,chang} and those where $\alpha $, $\beta $ and
$\delta $ are non-zero but $a$,  $b$ and $d$ vanish
\cite{benatti,liu,chang,klapdor}.
Benatti and Floreanini
\cite{benatti} consider a quite general model (all parameters
non-zero) initially, but later concentrate on the case when
$d=\beta =0$ and also $\delta =0$, although they impose the
condition of complete positivity.
They also consider all
parameters non-zero using a second order approximation.
Our models
below are all special cases of the models considered by Benatti
and Floreanini. Liu and collaborators \cite{liu} set $a=b=d=0$,
although in this case the oscillation probability can only be
computed numerically.
Here we either set both $a$ and $\alpha $
to be zero, or both are non-zero, so our probabilities do not
tally with theirs.
Chang et al \cite{chang} consider the same
model as \cite{liu}, but set $\beta =0$ in order to obtain an
analytic probability.
Again, our models do not fit into their picture.
They also consider the leading order corrections to
the standard oscillation probability for $a$, $b$, and $\alpha $
non-zero and $d$, $\beta $, and $\delta $ vanishing.
Ma and Hu
\cite{ma} work with the second of Chang et al's models, this time
with exact analytic results for the oscillation probability.
Models 1 - 4 below consider
various combinations of parameters within this model. Finally,
Klapdor-Kleingrothaus and collaborators \cite{klapdor} consider a
model in which only $\alpha $ and $\delta $ are non-zero.

We first consider a class of models in which
we set $d$ and $\beta $ equal to zero, so that
both the terms $M_{13}(E,L)$ and $M_{31}(E,L)$ in equation
(\ref{eq:genprob})  vanish.
The forms of the quantities $M_{11}(E,L)$ and $M_{33}(E,L)$
in equation (\ref{eq:genprob}) are
derived by solving the differential equations (\ref{diffeqn:rho})
describing the time evolution of the components of the density
matrix, using suitable initial conditions.
The answer (with $c=\hbar=1$) is:
\begin{eqnarray}
M_{11} & = & \left\{
\begin{array}{ll}
e^{-\Gamma _{{\rm {sum}}}} \cos \Gamma _{1}  & {\mbox {if $a=\alpha $;}} \\
e^{-\Gamma _{{\rm {sum}}}} \left(
\cos \Gamma _{2} + \frac {\Gamma _{{\rm {diff}}}}{\Gamma _{2}}
\sin \Gamma _{2}
\right)
& {\mbox {if $b=0$;}}
\end{array}
\right.
\nonumber
\\
M_{33} & = & e^{-2\delta L}  ; \label{eq:M11M33}
\end{eqnarray}
with
\begin{eqnarray}
\Gamma _{{\rm {sum}}} & = & \left( \alpha + a \right) L ;
\nonumber
\\
\Gamma _{{\rm {diff}}} & = & \left( \alpha - a \right) L;
\nonumber
\\
\Gamma _{1} & = &
2 \left[ \left(
 \frac {\Delta m^{2}L}{4E}   \right) ^{2} -b^{2}L^{2}
\right] ^{\frac {1}{2}} ;
\nonumber
\\
\Gamma _{2} & = &
2 \left[ \left(
 \frac {\Delta m^{2}L}{4E}   \right) ^{2}
 -\frac {1}{4} \left( \alpha -a \right) ^{2} L^{2}
\right] ^{\frac {1}{2}} . \label{eq:GAMMA}
\end{eqnarray}
Setting the quantum decoherence parameters in equations
(\ref{eq:M11M33}-\ref{eq:GAMMA}) equal to zero, we recover the
standard oscillation probability.
The other limit of interest is
if we set the standard oscillation parameter $\Delta m^{2}=0$ (but
keeping the usual mixing angle $\theta $ non-zero).
In this case
atmospheric neutrino oscillations would be due to quantum
decoherence effects only.
Given the widespread acceptance of the
standard oscillation picture, such a scenario may seem highly
unlikely, but the only existing analysis of experimental data
\cite{lisi,fogli} found that, while oscillations due to quantum
decoherence only are unfavoured by the experimental data, the
evidence is not overwhelming.
However, in this paper we shall mostly be
concerned with quantum decoherence effects as modifications of the
standard picture.
If we do set $\Delta m^{2}=0$ in equation
(\ref{eq:GAMMA}),
it is clear that we can obtain a sensible oscillation probability
only if $\Gamma _{1}$ and $\Gamma _{2}$ are real, which means that
$a=\alpha $ and $b=0$.
If, however, $\Gamma _{1}$
and $\Gamma _{2}$ are imaginary when $\Delta m^{2}=0$,
it is possible to write the oscillation probability
in terms of exponential (damping) terms only,
without any oscillations.
Given the success of the phenomenology
of neutrino oscillations, we do not consider this possibility
further.

The parameter $\delta $ corresponds to energy non-conserving
effects in the atmospheric neutrino system.
However, since the
terms in the probability (\ref{eq:genprob}) which contain $\delta
$ are multiplied by $\cos ^{2} (2\theta )$, and the current
experimental best-fit value for $\cos ^{2} (2\theta )$ is zero
\cite{fogli}, this energy non-conserving parameter has a
negligible effect on the oscillation probability, and so we will
not consider it further in this section.
However, see section
\ref{sec:exp} for bounds on this parameter from other experiments.
The models of this first type we shall consider are therefore:
\begin{enumerate}
\item
The simplest possible model is when $b=\delta =0$ and
$a=\alpha \neq 0$.
It is this model which has been studied for Super-Kamiokande
and K2K data \cite{lisi,fogli}, and the time evolution of the
density matrix satisfies the conditions of complete positivity and
energy conservation within the atmospheric neutrino system.
Furthermore, only in this model does the time-evolution of the density matrix
(\ref{dens:matnp}) have the Lindblad form (\ref{eq:lindblad}).
\item
We set $b=\delta =0$, with $a$ and $\alpha $ non-zero but not
equal. This is the simplest possible generalization of the
previous model; and energy is conserved in this case (as it is in
the next two models as well).
This model is probably the most
likely generalization to be of experimental significance, since
the consensus in the literature is that $b$ will be significantly
smaller than either $a$ or $\alpha $.
However, this model violates
the condition of complete positivity \cite{benatti}.
In this and the following two
models, oscillations cannot be accounted for solely by quantum
decoherence, and we must include a non-zero $\Delta
m^{2}$ in the probability.
\item
In our third model we consider only a non-zero $b$ in (\ref{eq:GAMMA}),
setting all other quantum decoherence parameters (including $a$
and $\alpha $) to vanish.
The form of the probability (\ref{eq:genprob}) is such that
the sign of $b$ cannot be measured, so we assume from henceforth that
$b$ is positive (equivalently, we are considering only $|b|$).
In the literature, the assumption is
frequently made that $b$ will be much smaller than either $a$ or
$\alpha $, which would rule this model out. However, we include it
here so that we can look specifically at the effect of $b$ on the
system.
Like the previous model, this one also violates complete
positivity \cite{benatti}.
\item
Our next model is, in effect, a
combination of the previous two.
We set $a=\alpha $ and $b$ to be
non-zero, but have $\delta $ and all other quantum decoherence
parameters vanishing.
Even this more general model does not
satisfy complete positivity \cite{benatti} as we have set $\delta =0$.
In studying this model in the next section, we find no
additional information on the parameter space of $\alpha $, $a$
and $b$ than we did by studying models 2 and 3, which implies that
it is sufficient to consider varying the quantum decoherence
parameters separately, at least as far as finding upper bounds on
sensitivity regions is concerned.
\end{enumerate}

Our second class of models violate energy conservation within the
atmospheric neutrino system.
We set $a=b=\alpha = \delta =0$ and
consider instead non-zero $d$ and $\beta $.
The oscillation probability will have the general form
(\ref{eq:genprob}).
Although, because the mixing angle $\theta $ for atmospheric
neutrino oscillations is such that $\cos ^{2} (2\theta )$ is very
close to zero, we are unable to measure $M_{33}$, $M_{31}$ or
$M_{13}$ directly, there are, in the two models below,
also modifications of $M_{11}$ which
we can probe, and so measure these other quantities indirectly.
This is in contrast to the situation in which the parameter
$\delta $ is non-zero, as this shows up only in $M_{33}$ if
$d=\beta =0$.
The models we consider are therefore:
\begin{enumerate}
\item[(5)]
Firstly, we set all parameters equal to zero except
$\beta$. The oscillation probability (\ref{eq:genprob}) in this
case is
\begin{eqnarray}
\label{eqn:probbeta}
\nonumber
P[\nu_{\mu}\rightarrow\nu_{\tau}]&=&\frac{1}{2}\left\{\cos^{2}2\theta
\left[1-\frac{\omega^{2}}{\Omega_{\beta}^{2}}+
\frac{\beta ^{2}}{\Omega_{\beta}^{2}}\cos(2\Omega_{\beta}L)\right]\right.\\
&
&\left.+\sin^{2}2\theta\left[1+\frac{\beta ^{2}}{\Omega_{\beta}^{2}}
-\frac{\omega^{2}}{\Omega_{\beta}^{2}}\cos(2\Omega_{\beta}L)\right]\right\}
\end{eqnarray}
where $\Omega_{\beta}=\sqrt{\omega^{2}-\beta ^{2}}$ and
$\omega=\frac{\Delta m^{2}}{4E}$.
In both this model and the next,
it is not possible for oscillations to be due to decoherence only,
as the argument of the $\cos$ term becomes imaginary if we set
$\Delta m^{2}=0$.
The probability (\ref{eqn:probbeta}) contains only $\beta ^{2}$ terms
and so we consider only $\beta >0$ (or, equivalently, $|\beta |$).
\item[(6)]
In this model, we set all the quantum
decoherence parameters to zero except $d$.
The probability of
oscillation (\ref{eq:genprob}) in this case has the form:
\begin{eqnarray}
\label{eqn:probc}
\nonumber
P[\nu_{\mu}\rightarrow\nu_{\tau}]&=&\frac{1}{2}\left\{\cos^{2}2\theta
\left[1-\frac{\omega^{2}}{\Omega_{d}^{2}}
+\frac{d^{2}}{\Omega_{d}^{2}}\cos(2\Omega_{d}L)\right]\right.\\
\nonumber & & \left.
+\sin^{2}2\theta\left[1+\frac{d^{2}}{\Omega_{d}^{2}}\cos(2\Omega_{d}L)
-\frac{\omega^{2}}{\Omega_{d}^{2}}\cos(2\Omega_{d}L)\right]\right.\\
& & \left. +\sin4\theta\left[\frac{d}{\Omega_{d}}\sin(2\Omega_{d}L)\right]\right\}
\end{eqnarray}
where $\Omega_{d}=\sqrt{\omega^{2}-d^{2}}$.
This is the only
model in which the quantities $M_{13}(E,L)$ and $M_{31}(E,L)$ are
non-zero.
Although both $d$ and $d^{2}$ appear in the probability (\ref{eqn:probc}),
in fact only $d^{2}$ can be measured with atmospheric neutrinos because
$\sin 4\theta \sim 0$.
Therefore we consider only $d>0$ (or $|d|$).
\end{enumerate}
We are now in a position to examine ANTARES sensitivity to these quantum decoherence
effects.

\section{ANTARES sensitivity to quantum decoherence in atmospheric
neutrino oscillations}
\label{sec:oscfit}

ANTARES is sensitive to high energy atmospheric neutrinos, with
energy above 10 GeV \cite{ANTARES}.
Although in some sense these
are a background to the neutrinos of astrophysical origin which
are the main study of a neutrino telescope, nevertheless the
detection of very high energy atmospheric neutrinos which have
very long path lengths (of the order of the radius of the Earth)
is complementary to other long baseline neutrino experiments such
as MINOS \cite{MINOS}, OPERA \cite{OPERA},
K2K \cite{K2K}, KamLAND \cite{KamLAND} and  CHOOZ \cite{CHOOZ}.
Our analysis of the sensitivity of ANTARES to quantum
decoherence effects in atmospheric neutrinos is based on a
modification of the analysis of standard atmospheric neutrino
oscillations with ANTARES \cite{ANTARES,antosc}.
We would anticipate that
other high energy neutrino telescopes such as AMANDA \cite{AMANDA}
and ICECUBE \cite{ICECUBE} would also be able to probe these
effects.

\subsection{Simulations and analysis}
\label{sec:simulation}

Our simulations are based on previous ANTARES simulations of the sensitivity
to standard neutrino oscillations \cite{antosc}.
Further details of the ANTARES detector, detector simulation and event signals
can be found in \cite{antosc}.
Here we summarize the main features of the simulations which are pertinent to
our discussion.
Atmospheric neutrinos events were generated using Monte-Carlo production.
A spectrum proportional to $E^{-2}$ for the neutrinos was assumed,  for
energies in the range 10 GeV $< E <$ 100 TeV.
The zenith angle distribution is assumed to be isotropic.
Twenty-five years of data were simulated, so that errors from the MC statistics
could safely be ignored.
Event weights were used to adapt the MC flux to a real atmospheric neutrino flux.
We used the Bartol theoretical flux \cite{bartol}, although the results will not be
significantly changed by using different theoretical fluxes.
There may be a more significant change in our results if the spectral index is modified.

Our simulations are based on three years' data taking with ANTARES.
This corresponds to roughly 10000 atmospheric neutrino events.
All errors are purely statistical, with simple Gaussian errors assumed.
By varying the form of the oscillation probability to include quantum
decoherence effects, spectra in either $E$, $E/L$ or $L$ are produced.
Some examples of these are discussed in the next subsection.
Since ANTARES is particularly sensitive to the zenith angle $\vartheta $ and
thereby the path length $L$, the spectra in $E/\cos \vartheta $ are used in the sensitivity
analysis.

To produce sensitivity regions, we used a $\chi ^{2}$ technique,
comparing the $\chi ^{2}$ for oscillations with decoherence (or decoherence alone)
with that for the no-oscillation hypothesis.
The total normalization is left as a free parameter in the sensitivity analysis, so that
our results in section \ref{sec:regions}
are not affected by the normalization of the atmospheric neutrino flux.
The sensitivity regions at both 90 and 99\% confidence level are produced.

Even though our analysis is independent of the normalization of the atmospheric neutrino
flux, in order to produce spectra based on numbers of events, as in the following
subsection, some normalization of the total flux is necessary.
For standard oscillations without quantum decoherence, bins with high $E/\cos \vartheta $
can be used to normalize the flux as standard oscillations are negligible at sufficiently
high $E/L$ (as can be seen in figure \ref{fig:standard1}).
This method can also be applied for quantum decoherence models in which the decoherence
parameters are inversely proportional to the neutrino energy, so that decoherence also becomes
negligible at high energy.
However, when the decoherence parameters are proportional to the neutrino energy squared,
decoherence is significant at high energies but negligible at low $E/L$.
Therefore, in this latter case, we may use the low $E/\cos \vartheta $ bins
to normalize the flux.

We have studied six different models of quantum decoherence, as
outlined in the previous section, and for each model there are
three different ways in which the quantum decoherence
parameters may depend on the neutrino energy.
We shall not discuss
the details of all eighteen of our simulations, but instead
summarize our results and discuss a few models in more detail to
illustrate the essential features.
In our simulations, for
practical reasons, we had at most three varying parameters, one of
which was always the mixing angle $\theta $ which facilitated
comparisons with previous work \cite{lisi,fogli,antosc}.
In models with
only one non-zero decoherence parameter (namely models 1, 3, 5 and
6 above) we also varied $\Delta m^{2}$, to check that our
sensitivity regions included the current best-fit values
\cite{lisi,fogli}.
In models with two non-zero
decoherence parameters (models 2 and 4), we fixed $\Delta
m^{2}=2.6\times10^{-3}$ eV$^{2}$.
The simulations then give us a
three-dimensional region of sensitivity at 90 and 99 percent
confidence level, and we plotted the boundaries of the projections of the surface
bounding this volume onto the co-ordinate planes, where necessary
also plotting the surface itself.

\subsection{Spectra}
\label{sec:spectra}

As well as data based on total numbers of events, the spectra of these events
in (typically) $E/L$ is also important in neutrino oscillation physics.
For atmospheric neutrinos, Super-Kamiokande \cite{SKspectra} has reported
an oscillatory signature in the $E/L$ spectrum of events, and similar
spectral analyses have been performed for KAMLAND \cite{KAMLANDspectra} and
K2K \cite{K2Kspectra}.
As well as being used in our analysis of sensitivity regions, the spectra themselves
are important for searching for possible decoherence effects and also for bounding
the size of such effects.
The overall normalization of the flux, required to produce these spectra, can be found
using either the high or low $E/\cos \vartheta $ bins, depending on the situation,
as described in the previous subsection.

\begin{figure}
\begin{center}
\includegraphics[angle=270,width=10cm]{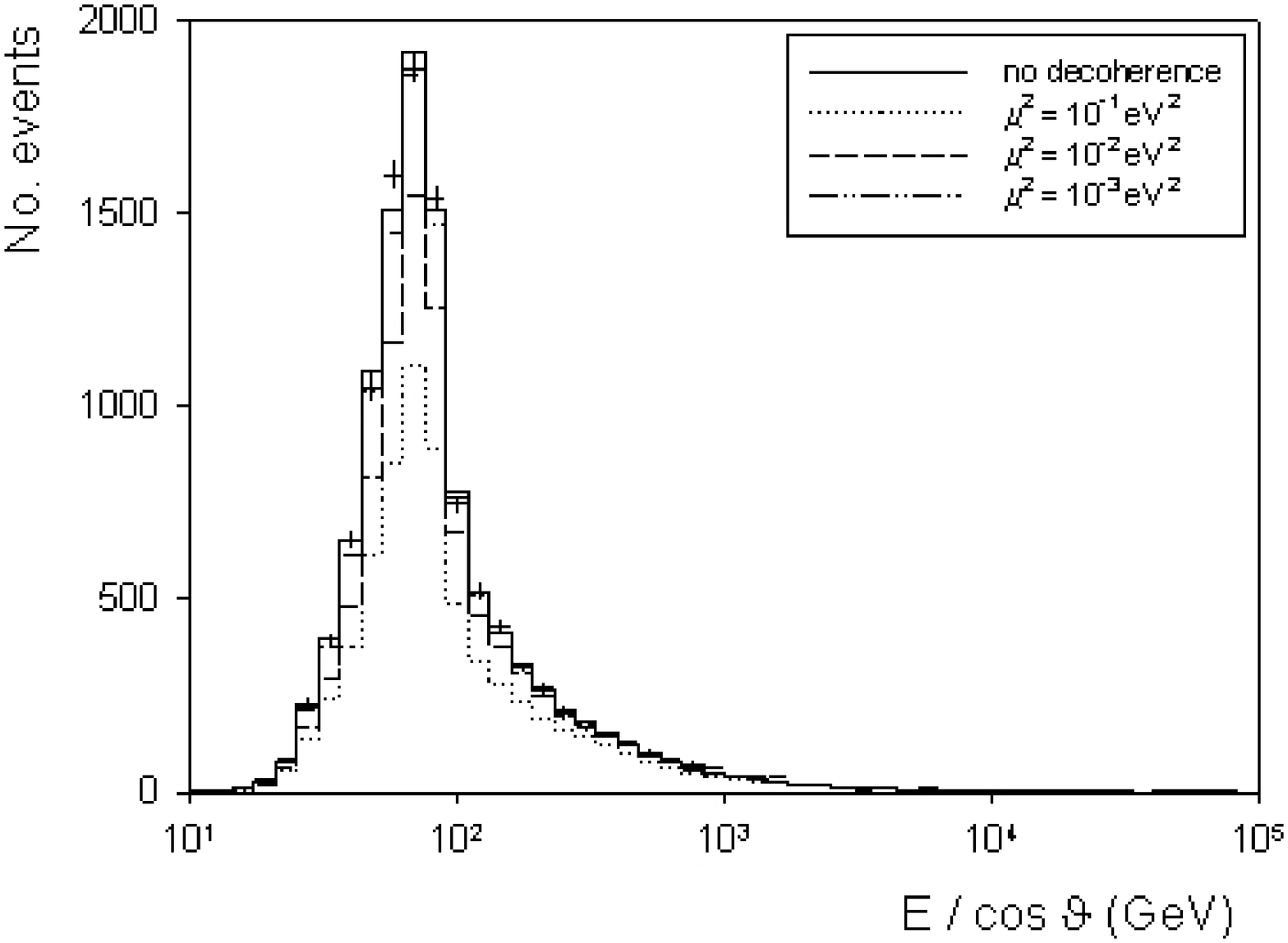}\\
\includegraphics[angle=270,width=10cm]{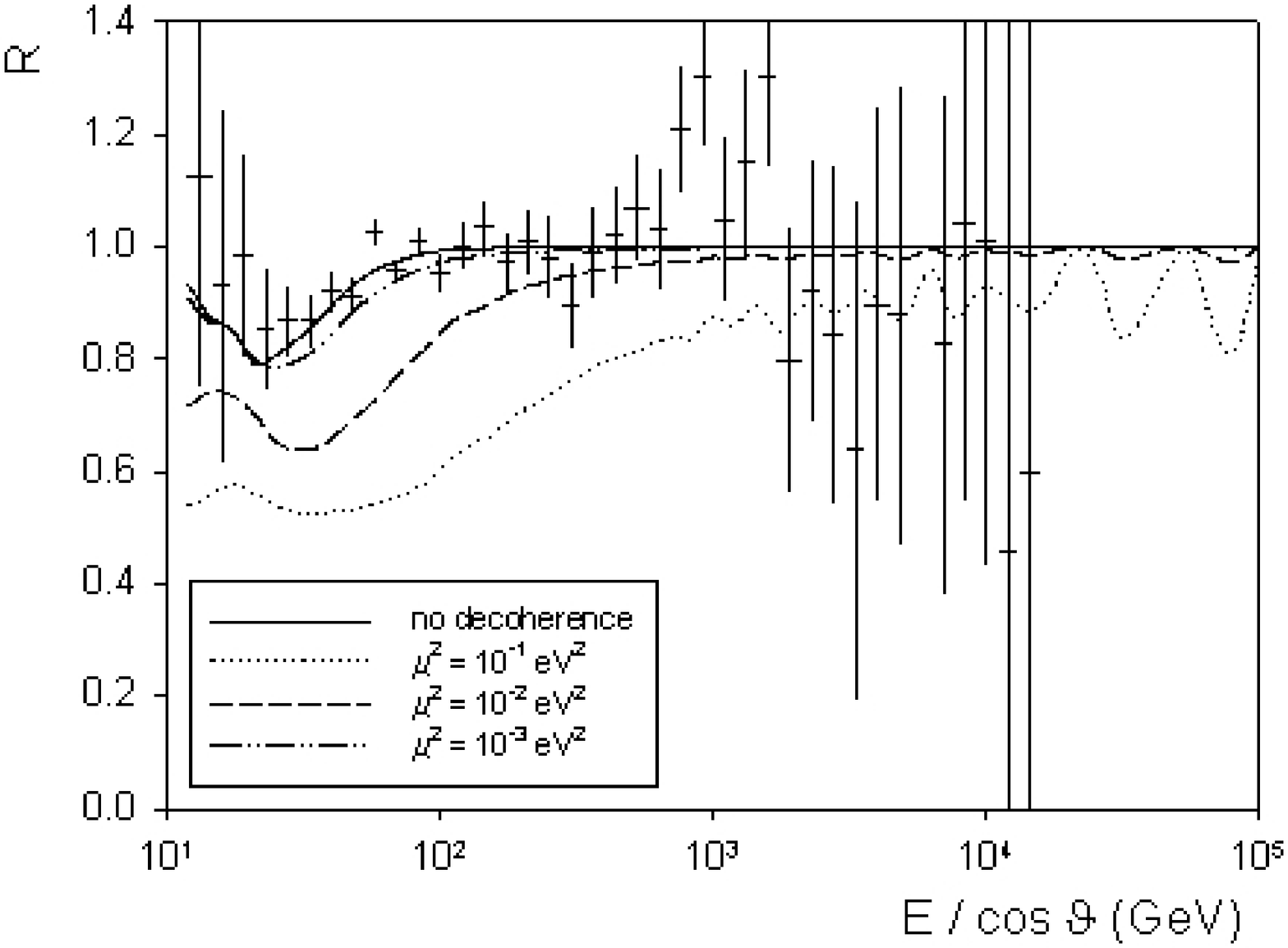}
\caption{Spectrum of events as a function of $E/\cos \vartheta $ (top) and
ratio of the number of events compared to no oscillations (bottom).
In each graph, the solid line represents MC simulation of
standard oscillations without decoherence, and the
dotted lines represent MC simulation of
standard oscillations plus decoherence.
The data points correspond to a three-year measurement, assuming standard oscillations
only without decoherence.
The decoherence parameters are inversely proportional to the neutrino energy.}
\label{fig:spectra1}
\end{center}
\end{figure}

\begin{figure}
\begin{center}
\includegraphics[angle=270,width=10cm]{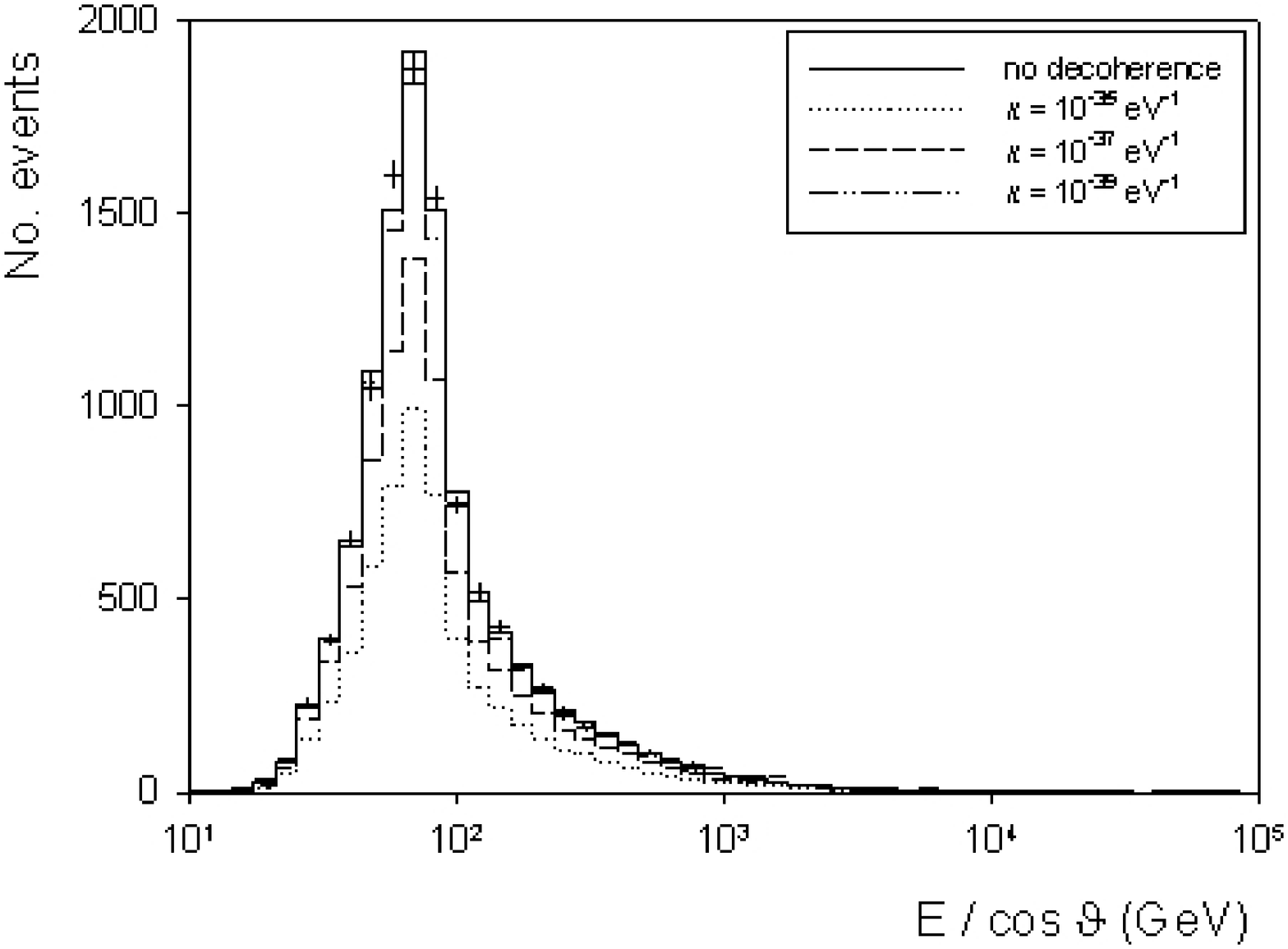}\\
\includegraphics[angle=270,width=10cm]{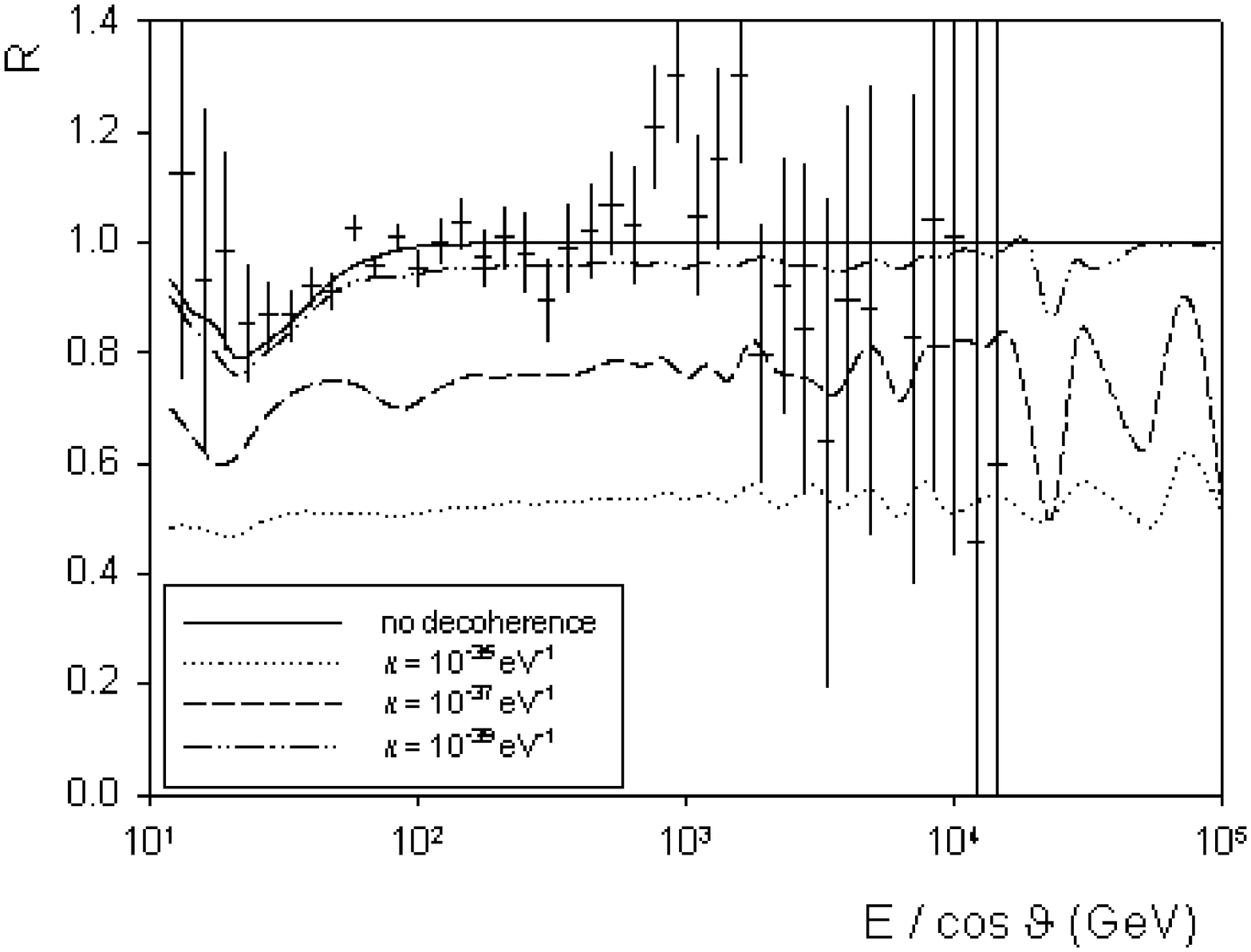}
\caption{Spectra of events as in figure \ref{fig:spectra1} but for the case when the
decoherence parameters are proportional to the neutrino energy squared.}
\label{fig:spectra2}
\end{center}
\end{figure}

Figures \ref{fig:spectra1}, \ref{fig:spectra2} show typical spectra of events
as a function of $E/\cos \vartheta $ (corresponding
to $E/L$), where $E$ is the reconstructed energy and $\vartheta $
the zenith angle, not to be confused with the mixing angle $\theta $.
In each case, the solid line is the MC simulation with standard oscillations and no
decoherence, using the values of the oscillation parameters in (\ref{eq:bestfit}),
while the dotted and dashed lines are the MC simulation with standard oscillations
and decoherence together.
The spectra in figure \ref{fig:spectra1} are for the model in which the quantum
decoherence parameters are inversely proportional to the neutrino energy,
while in figure \ref{fig:spectra2} we consider quantum decoherence parameters which
are proportional to the neutrino energy squared.
In both figures \ref{fig:spectra1}, \ref{fig:spectra2}, we have also plotted the ratio
of the number of events compared with no oscillations.
It can be seen that in both cases
quantum decoherence results in a significant reduction in the number of
events.
However, there is also a difference in the shape of the spectra.
In both models, when the decoherence parameters are comparatively large,
the spectrum in the ratio of the number of events (lower plots) is
much flatter in the region below about 150 GeV,
compared with standard oscillations (when there is
a `dip' in the ratio of the number of events).
For the case in which the decoherence parameters are inversely proportional to the
neutrino energy, the ratio of the number of events rises to one at very high energies,
but more slowly than the case of oscillations only.
However, it should be stressed that the values of the parameter $\mu ^{2}$
for which this is most noticeable in figure \ref{fig:spectra1} are larger than the current
best upper bound from SK and K2K data \cite{fogli}.
When the decoherence parameters are proportional to the neutrino energy squared
(figure \ref{fig:spectra2}), the spectrum of the ratio of the number of events is
extremely flat.

\begin{figure}
\begin{center}
\includegraphics[angle=270,width=10cm]{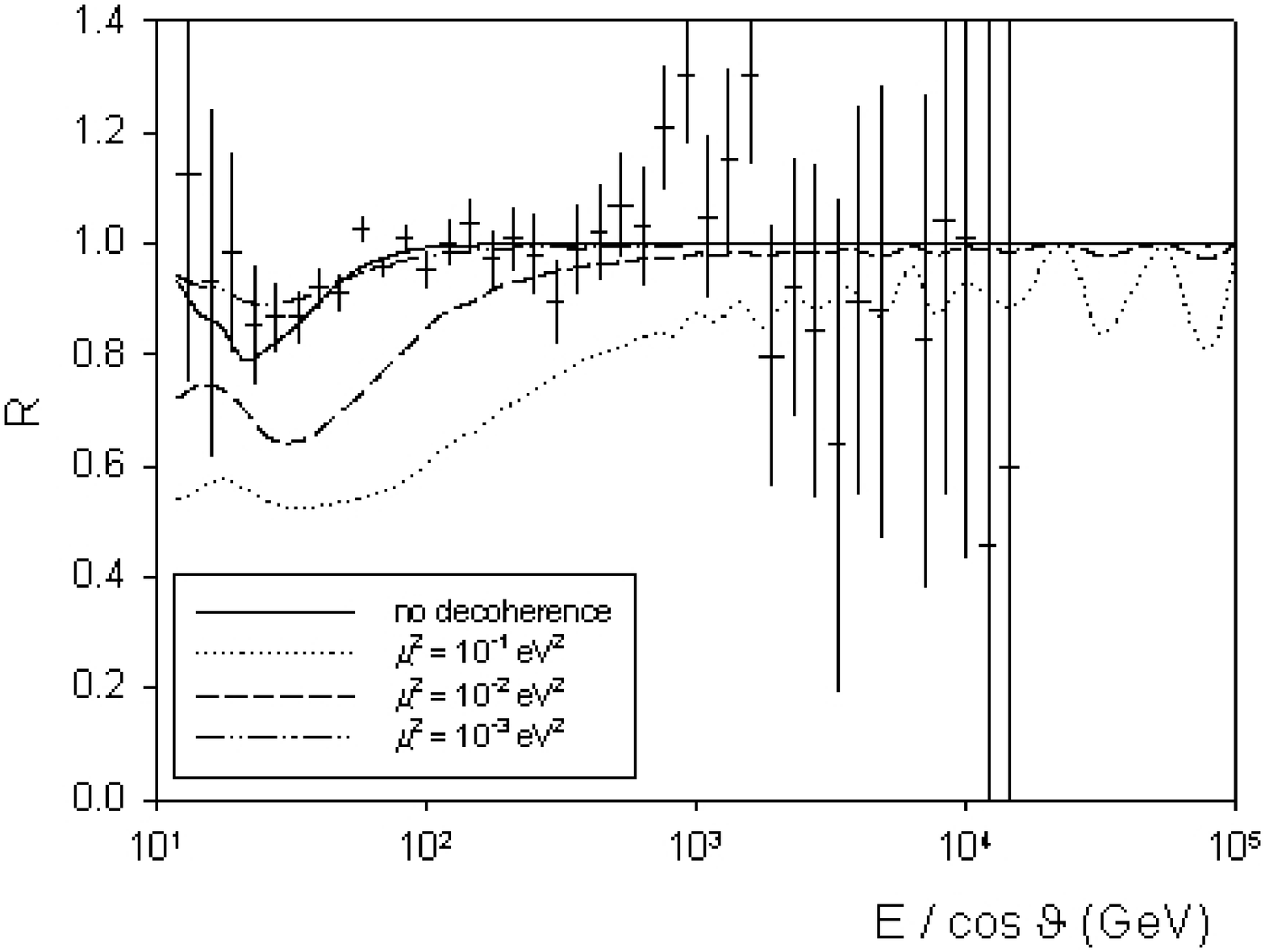}\\
\includegraphics[angle=270,width=10cm]{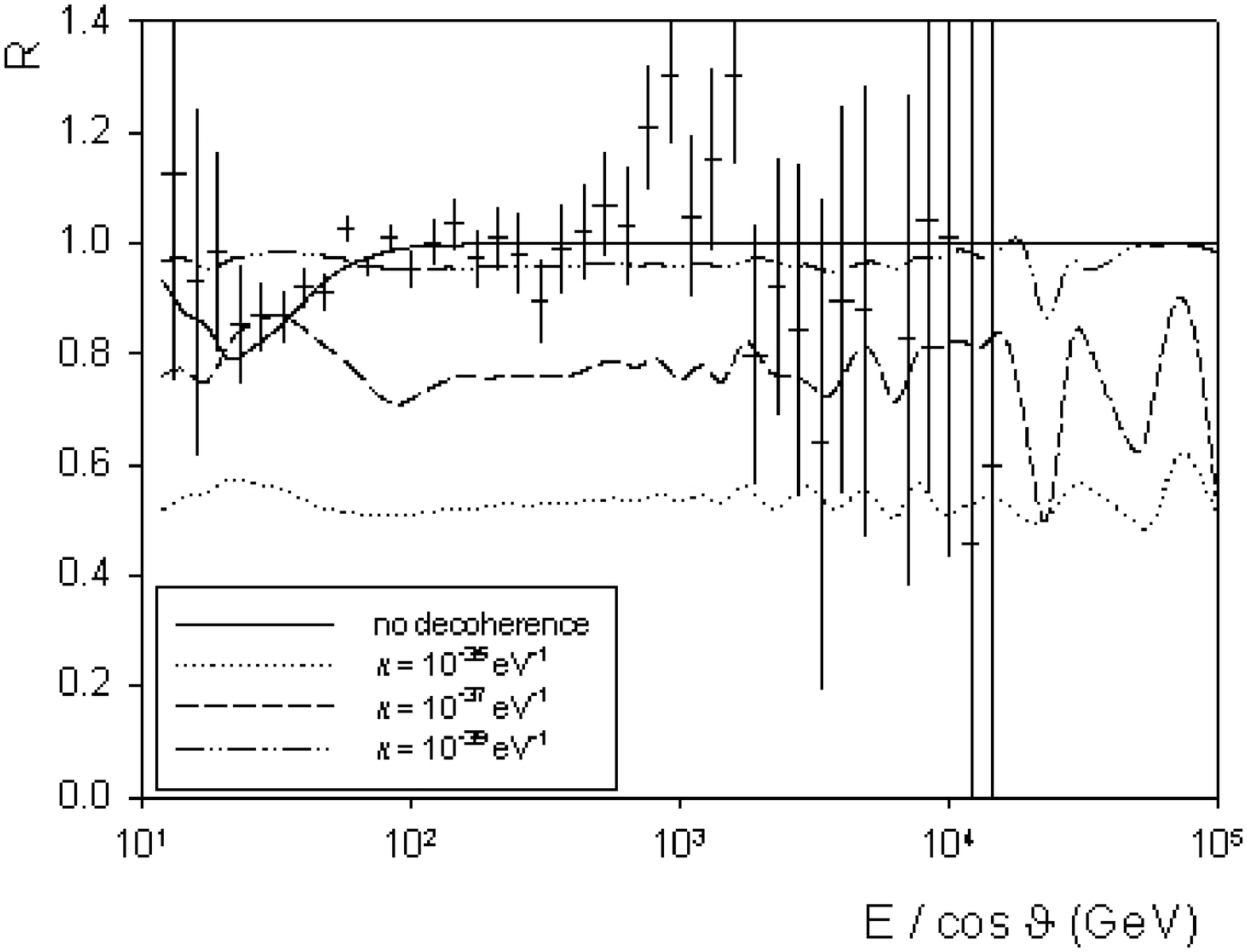}
\caption{Ratios of the number of events compared to no oscillations, as functions
of $E/\cos \vartheta $.
The solid line is for standard oscillations only without decoherence, for comparison.
The remaining curves are for no standard oscillations, and quantum decoherence only.
The upper plot is for quantum decoherence parameters inversely proportional to the
neutrino energy, and the lower plot for the case when the quantum decoherence
parameters are proportional to the neutrino energy squared.}
\label{fig:spectra3}
\end{center}
\end{figure}

We can also consider the situation in which there are no standard oscillations,
in other words, the squared mass difference $\Delta m^{2}$ is zero.
In this case, neutrino oscillations would be described entirely by decoherence.
In this case the ratios of the numbers of events compared
with the no-oscillation hypothesis are shown in figure \ref{fig:spectra3},
for decoherence parameters inversely proportional to the neutrino energy (upper
plot) and proportional to the neutrino energy squared (lower plot).
We have plotted in figure \ref{fig:spectra3} the ratio for the standard oscillations
only case for comparison (the solid line), but the other curves are for
pure decoherence without standard oscillations.
The difference in the spectra at $E/\cos \vartheta $ below about 150 GeV
is clear for the case when the decoherence parameters are proportional
to the neutrino energy squared,
with the decoherence-only model generally producing a peak in the
spectrum where standard oscillations without decoherence has a dip.
When the decoherence parameters are inversely proportional to the neutrino
energy, the spectra look more like those for standard oscillations
plus decoherence (figure \ref{fig:spectra1}), except for small values of
$\mu ^{2}$, when the spectrum is flatter than for standard oscillations only.

It should be emphasised that the values of the quantum decoherence parameter $\kappa $
used to generate the spectra in figures \ref{fig:spectra2}
and \ref{fig:spectra3} are
many orders of magnitude smaller than the current experimental upper bound of
$10^{-19}$ eV${}^{-1}$ \cite{lisi}.
As the decoherence parameter $\kappa $ increases, the number of events decreases and
so it is clear from figures \ref{fig:spectra2} and \ref{fig:spectra3} that ANTARES
will be able to rule out even very small values of $\kappa $.
This can be understood from the oscillation probabilities as shown in the figures in section
\ref{sec:atmqg}.
For high energy neutrinos, the standard oscillation probability is negligible and so
the number of events depends only on the atmospheric neutrino flux.
However, with quantum decoherence, the oscillation probability becomes close to one-half
for atmospheric neutrinos and so a large fraction of the incoming muon neutrinos
will oscillate, resulting in a significant decrease in the number of
events seen.
For low energy neutrinos, as observed by current atmospheric neutrino experiments,
in the model we have concentrated on in this subsection, the effects of quantum decoherence
are negligible with these values of the quantum decoherence parameters because the
effects are proportional to the neutrino energy squared.

This is the main result of our paper: ANTARES measures high-energy neutrinos and so
will be able to put stringent bounds on quantum decoherence effects which are proportional
to the neutrino energy squared.

\subsection{Sensitivity regions}
\label{sec:regions}

We now turn to a discussion of the ANTARES sensitivity regions from our simulations.
We discuss the simplest model in some detail as it reveals most of the salient features.
The remaining models are summarized briefly.
We are particularly interested in the upper bounds on the sensitivity regions.
We focus mainly on the case where the quantum decoherence parameters are proportional
to the neutrino energy squared, as here we will have the most significant improvement on
current experimental bounds.

\subsubsection{Model 1}
\label{sec:one}

This is the model in which $\alpha = a$ and has been studied for Super-Kamiokande and K2K data
\cite{lisi,fogli}.
They found the best fit to the data when the quantum decoherence parameter
is inversely proportional to the neutrino energy \cite{lisi}.
A detailed analysis \cite{fogli} of this case then found that the favoured
situation was standard oscillations without decoherence, but that the data
could not entirely rule out oscillations due to quantum decoherence only.
We show our results for this model with the parameters $\alpha =a$ inversely
proportional to the neutrino energy in detail for comparison purposes,
but our main results are concerned with the case in which
the parameters are proportional to the neutrino energy squared, as it is in this
latter case that we can significantly improve current bounds.

\begin{figure}
\begin{center}
\includegraphics[angle=270,width=8cm]{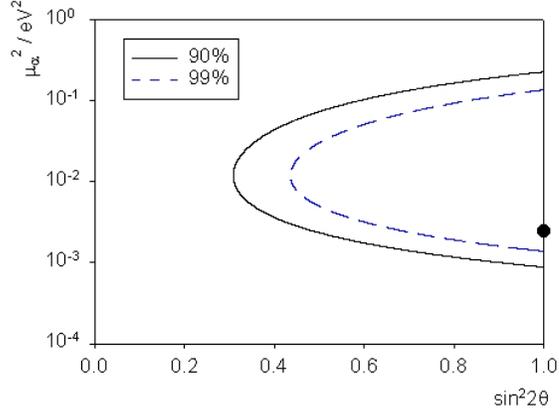}
\end{center}
\caption{Sensitivity contours at 90 and 99 percent confidence
level for quantum decoherence effects only (no standard
oscillations), with parameter $\mu_{\alpha} ^{2}$. The circle
denotes the value $\mu_{\alpha} ^{2}=2.44 \times 10^{-3}$
eV${}^{2}$, which is the best fit to combined Super-Kamiokande and
K2K data \cite{fogli}.}
\label{fig:skonlyexc}
\end{figure}

\begin{figure}
\begin{center}
\includegraphics[angle=270,width=8cm]{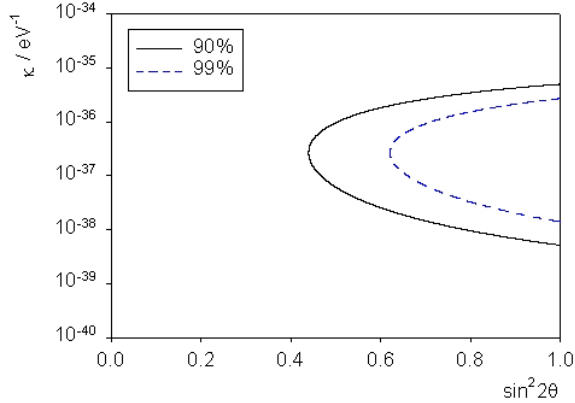}
\end{center}
\caption{Sensitivity contours at 90 and 99 percent confidence
level for quantum decoherence effects only (no standard
oscillations), with parameter $\kappa _{\alpha}$.}
\label{fig:kappaonlyexc}
\end{figure}

Firstly, consider the case in which $\Delta m^{2}=0$, so that there are no standard
oscillations and only quantum decoherence.
Figure \ref{fig:skonlyexc} shows the sensitivity curves in this case, for
$\alpha =a$ inversely proportional to the neutrino energy, and
figure \ref{fig:kappaonlyexc} for $\alpha = a$ proportional to the neutrino energy
squared.
It is somewhat surprising that current experimental data
\cite{fogli} only slightly disfavours this model, in comparison to
the standard oscillation picture.
The best fit value for combined
Super-Kamiokande and K2K data is $\mu_{\alpha} ^{2} = 2.44 \times
10^{-3}$ eV${}^{2}$ with $\sin ^{2} (2\theta )=1.00$ \cite{fogli},
which lies within our region of sensitivity, and is shown by a
solid circle in figure \ref{fig:skonlyexc}.
However, the K2K data
taken alone has a best fit of $\mu_{\alpha} ^{2}=2.46 \times
10^{-3}$ eV${}^{2}$, with $\sin ^{2} (2\theta )=0.86$ \cite{fogli}.
It is clear
from figure \ref{fig:skonlyexc} that the smallest values of
$\mu_{\alpha} ^{2}$ can be probed when $\sin ^{2} (2 \theta )$ is
closest to one, which is the region of parameter space in which we
are most interested.
It is also interesting to note the
similarity in the value of $\mu_{\alpha}^{2}$ to that of $\Delta
m^{2}$ for standard oscillations.

For the case in which $\alpha =a $ are proportional to the neutrino energy squared,
it is important to note that the range of values of the parameter $\kappa _{\alpha }$
shown in figure \ref{fig:kappaonlyexc} is many orders of magnitude lower than
the current upper bound of $10^{-19}$ eV${}^{-1}$ \cite{lisi}.
Therefore ANTARES will be able to probe regions of parameter space currently
inaccessible to lower-energy experiments, and improve significantly on
existing experimental bounds.

We also studied pure decoherence for the case in which $\alpha =a$ do not depend
on the neutrino energy.
The sensitivity curves in this case have a very similar shape to those in
figures \ref{fig:skonlyexc} and \ref{fig:kappaonlyexc}, with sensitivity
to values of the parameter $\gamma _{\alpha }$ in the region $10^{-13}$ -- $10^{-14}$ eV,
which is a similar range to the current upper bound from SK data of $10^{-14}$ eV \cite{lisi}.

\begin{figure}
\begin{center}
\includegraphics[angle=270,width=8cm]{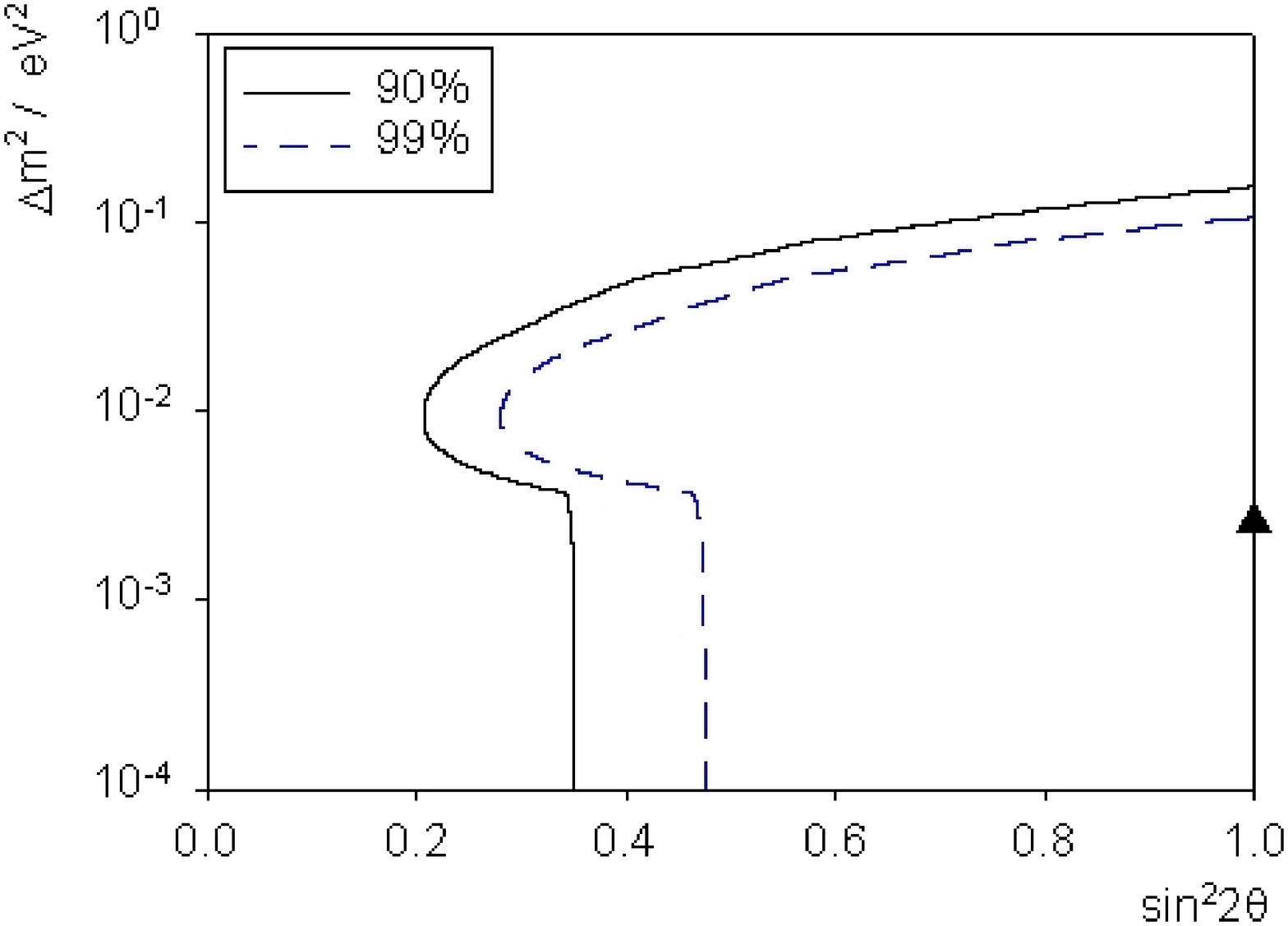}  \\
\includegraphics[angle=270,width=8cm]{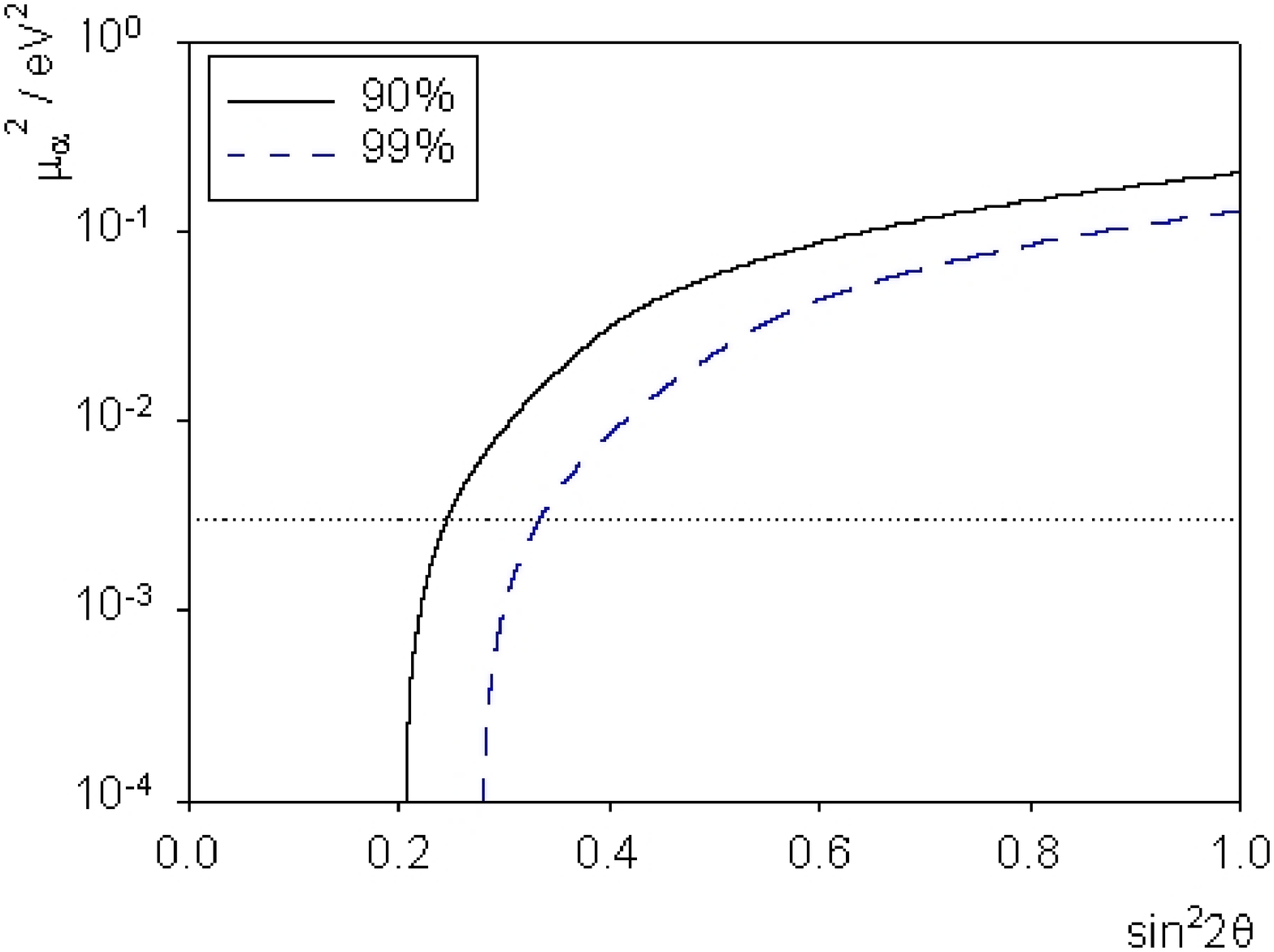} \\
\includegraphics[angle=270,width=8cm]{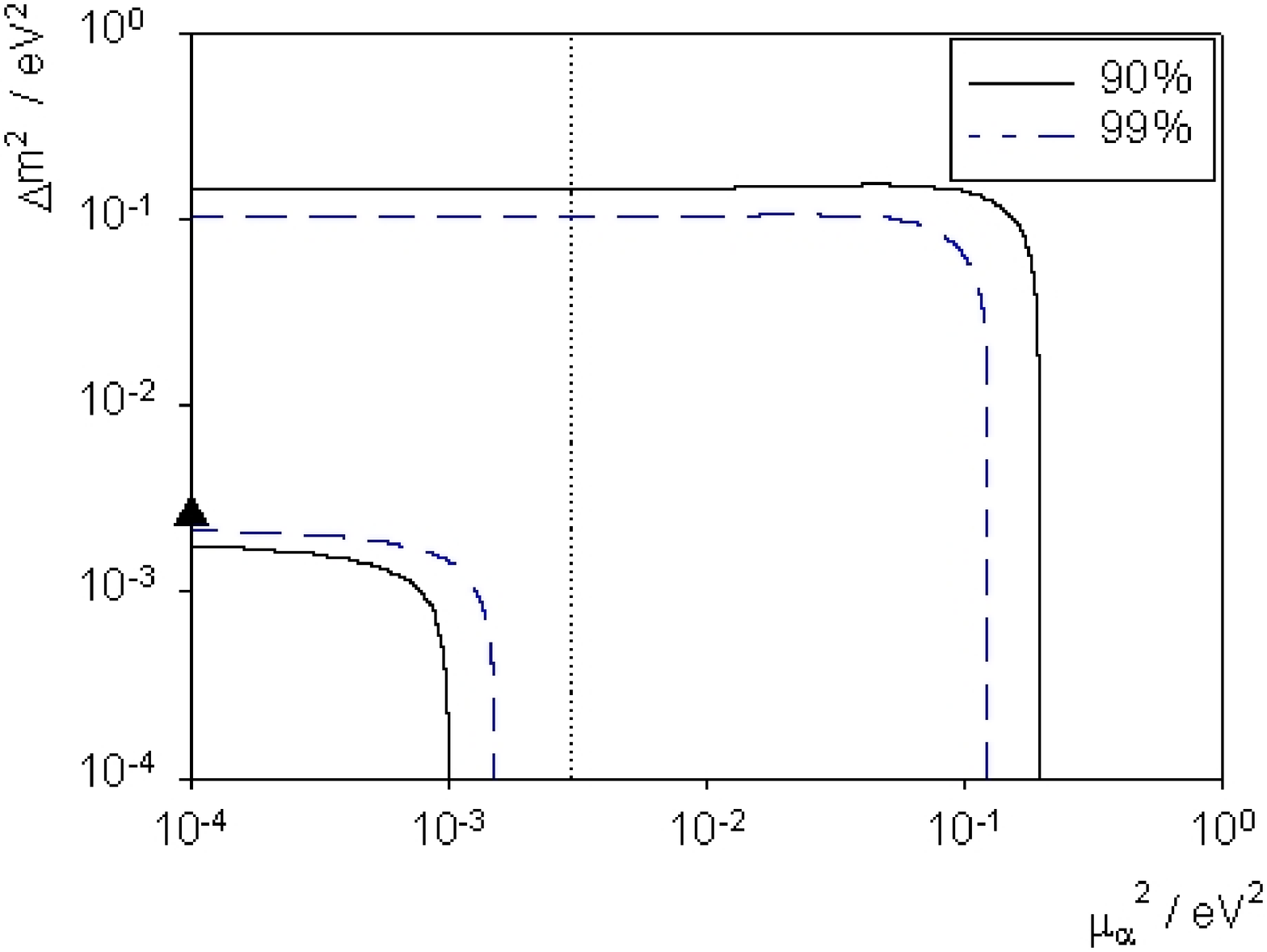}
\end{center}
\caption{Sensitivity contours at 90 and 99 percent confidence
level, when both oscillation and quantum decoherence parameters
($\mu_{\alpha}^{2}$) are non-zero.
The dotted line denotes the bound $\mu_{\alpha} ^{2} < 3
\times 10^{-3}$ eV${}^{2}$ from Super-Kamiokande and K2K data
\cite{fogli}, and the triangle the best fit value $\Delta
m^{2}=2.6 \times 10^{-3}$ eV${}^{2}$, $\sin ^{2}(2\theta )=1$
\cite{fogli}.}
\label{fig:skbothexc}
\end{figure}

\begin{figure}
\begin{center}
\includegraphics[angle=270,width=8cm]{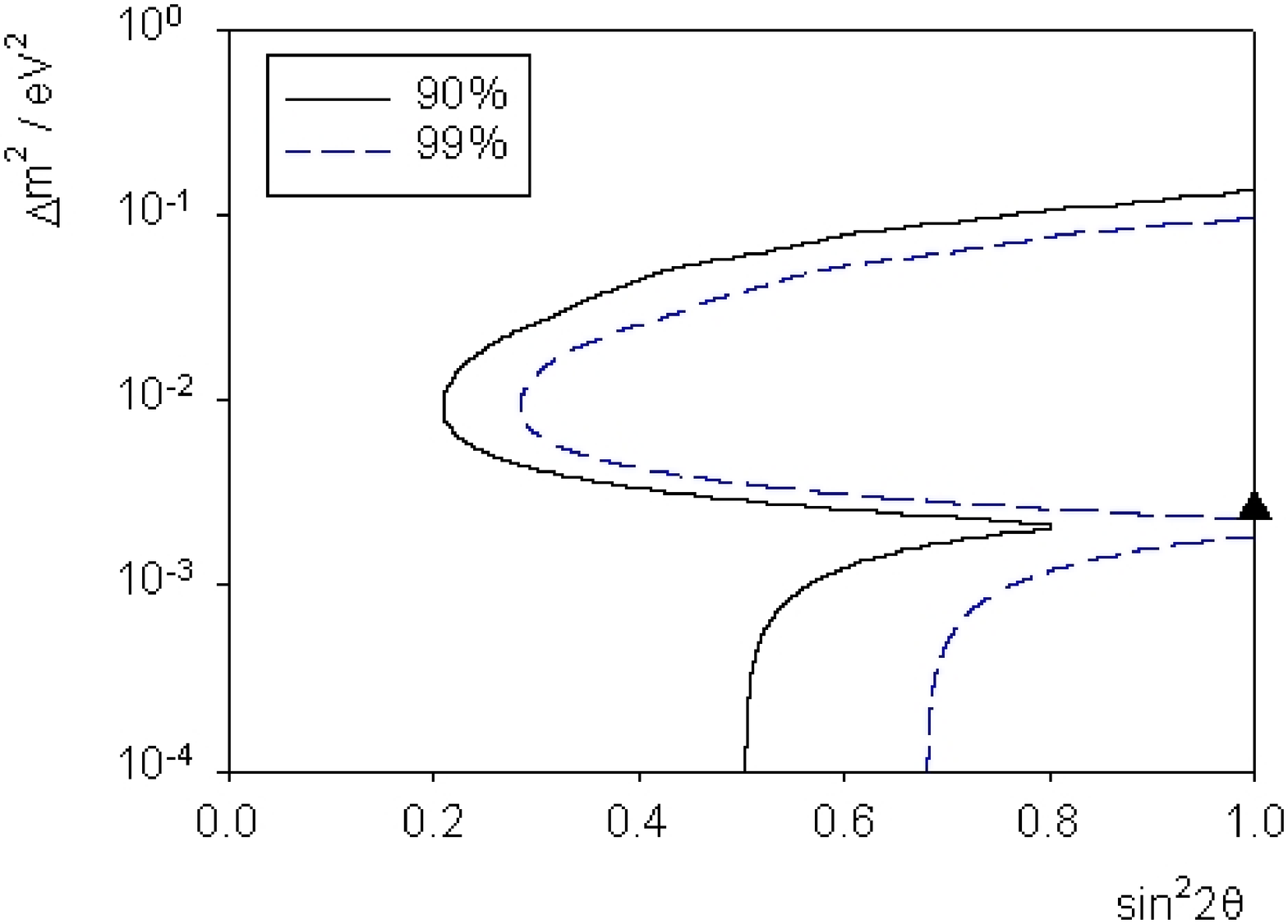}  \\
\includegraphics[angle=270,width=8cm]{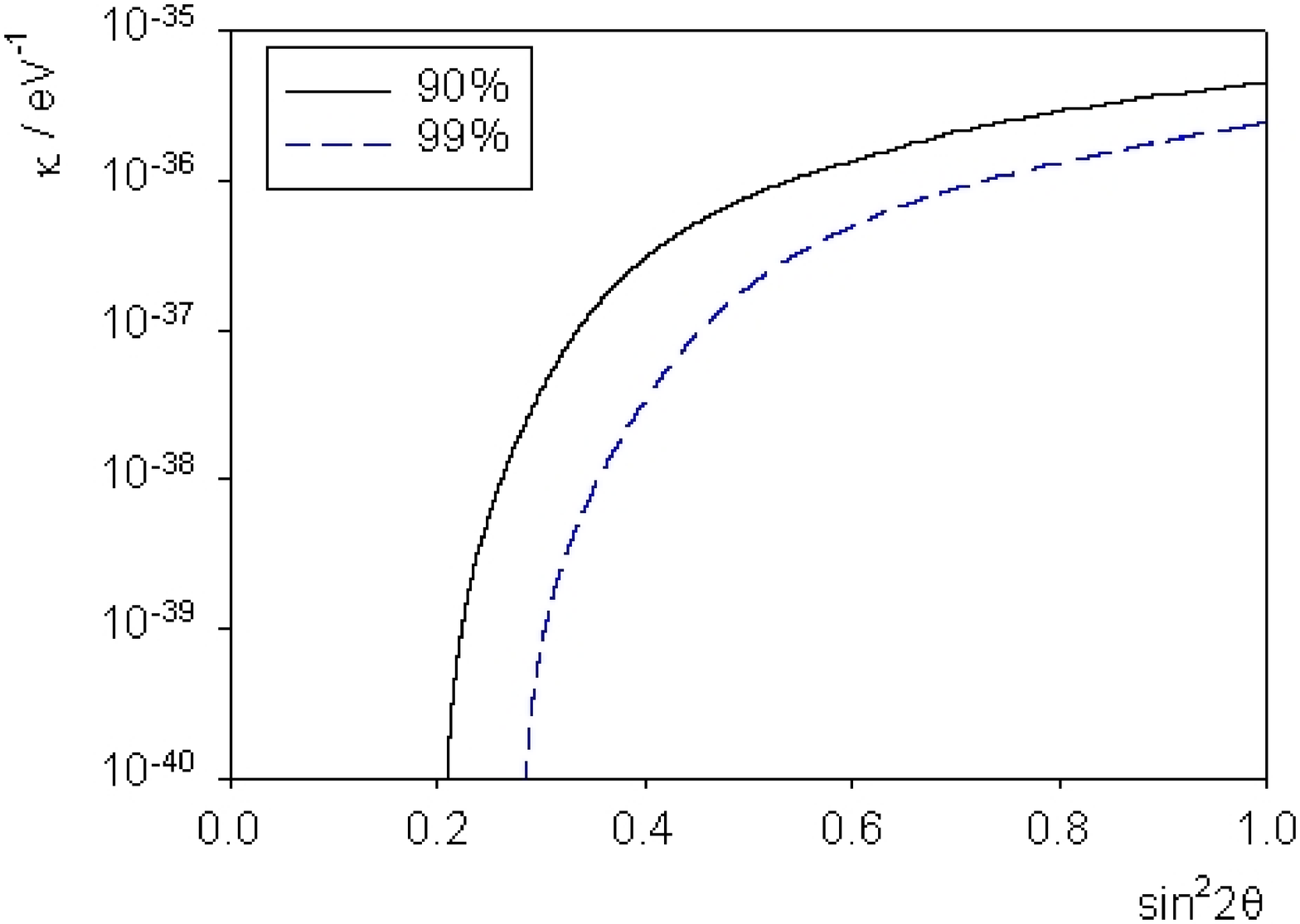} \\
\includegraphics[angle=270,width=8cm]{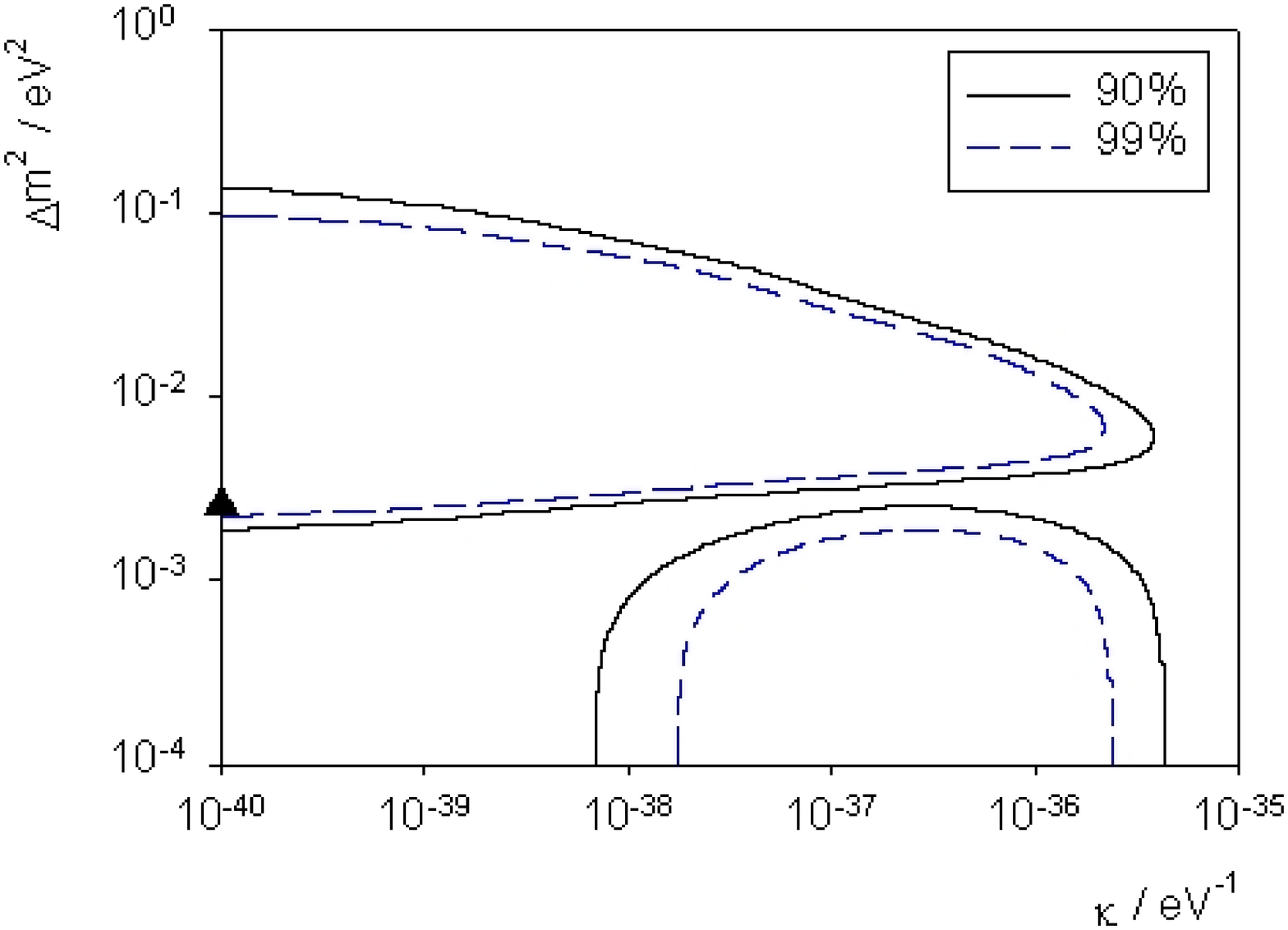}
\end{center}
\caption{Sensitivity contours at 90 and 99 percent confidence
level, when both oscillation and quantum decoherence parameters
($\kappa _{\alpha}$) are non-zero.
The triangle denotes the best fit value $\Delta
m^{2}=2.6 \times 10^{-3}$ eV${}^{2}$, $\sin ^{2}(2\theta )=1$
\cite{fogli}.}
\label{fig:kappabothexc}
\end{figure}

We also found sensitivity curves for the more general case when
both $\Delta m^{2}$ and $\mu_{\alpha}^{2}$, $\kappa _{\alpha }$
or $\gamma _{\alpha }$ (as applicable) are non-zero.
The
results we obtained are shown in figures \ref{fig:skbothexc}
and \ref{fig:kappabothexc} for the $\mu _{\alpha }^{2}$ and
$\kappa _{\alpha }$ cases, respectively.
We do not show the graphs for the $\gamma _{\alpha }$ case because they
are qualitatively very similar to the $\mu _{\alpha }^{2}$ case.
In figures \ref{fig:skbothexc} and \ref{fig:kappabothexc},
we have plotted projections of a three-dimensional volume onto
each of the co-ordinate planes, and the contours represent the
boundaries of the projected regions.
The sensitivity contours should be compared with the combined
Super-Kamiokande and K2K results \cite{fogli}.
In each figure, the
best fit values of $\Delta m^{2}=2.6 \times 10^{-3}$ eV${}^{2}$
and $\sin ^{2}(2\theta )=1$ \cite{fogli} are denoted by a
triangle, while the upper bound $\mu_{\alpha} ^{2}<3 \times
10^{-3}$ eV${}^{2}$ \cite{fogli} is marked by a dotted line
in figure \ref{fig:skbothexc}.
Our regions have a very similar shape to those constructed from data
in \cite{fogli}.

For the situation in which the quantum decoherence parameters are
inversely proportional to the neutrino energy (figure \ref{fig:skbothexc}),
it should be noted that the upper bound on the sensitivity region
at $\sin ^{2}(2\theta )=1$ is considerably higher than the upper
bound from data \cite{fogli}.
This is to be expected since ANTARES
measures only high energy atmospheric neutrinos (energy greater
than 10 GeV compared with a few GeV for Super-Kamiokande \cite{SK} and K2K
\cite{K2K}) and so is likely to be comparatively insensitive to
effects inversely proportional to the neutrino energy.
However, when the decoherence parameters are proportional
to the neutrino energy squared (figure \ref{fig:kappabothexc}),
the upper bound on the sensitivity region, $\kappa _{\alpha }< 10^{-35}$
eV ${}^{-1}$, is many orders of magnitude smaller than the corresponding bound
from Super-Kamiokande data of $10^{-19}$ eV${}^{-1}$ \cite{lisi}.
In the final case, when the quantum decoherence parameters are independent
of the neutrino energy, we find an upper bound on the sensitivity region of
$\gamma _{\alpha }<10^{-14}$ eV,
which is of a similar order of magnitude to the bound from Super-Kamiokande data
of $10^{-14}$ eV \cite{lisi}.

\subsection{Models 2, 3 and 4}
\label{sec:2-4}

These models involve combinations of the quantum decoherence
parameters $a$, $\alpha $ and $b$.
The upper bounds of the
sensitivity regions that we find for a particular parameter are essentially independent
of the particular combination of other parameters chosen to be
non-zero, and whether or not we fix $\Delta m^{2}$.
The parameters $a$ and $\alpha $ have the same sensitivity regions when they are
not necessarily equal, and whenever the parameter $b$ is included,
it has a upper bound corresponding to the value at which the
factors $\Gamma _{1}$ and $\Gamma _{2}$ (\ref{eq:GAMMA}) (as
applicable) become imaginary.

\begin{figure}
\begin{center}
\includegraphics[angle=270,width=8cm]{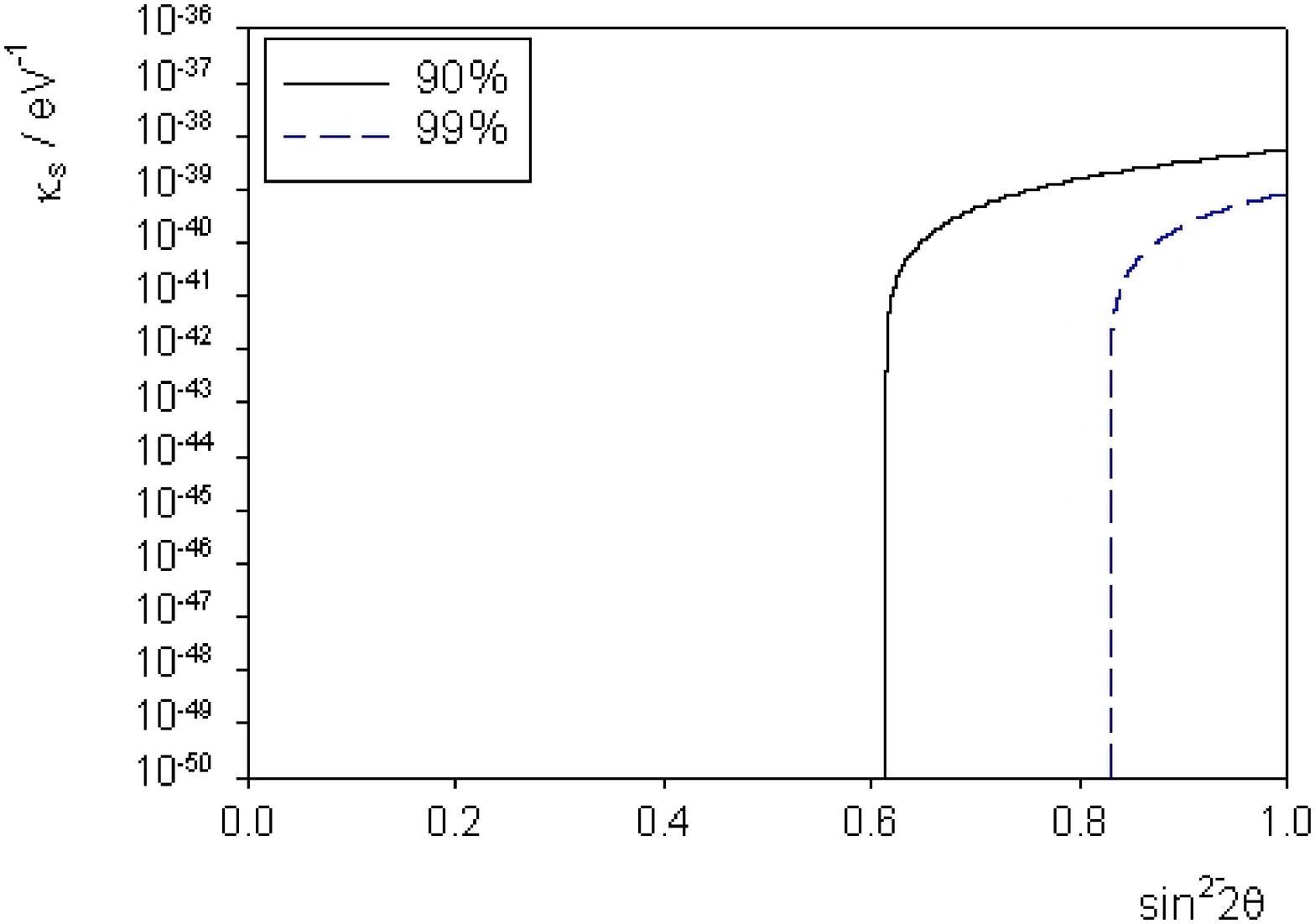}\\
\includegraphics[angle=270,width=8cm]{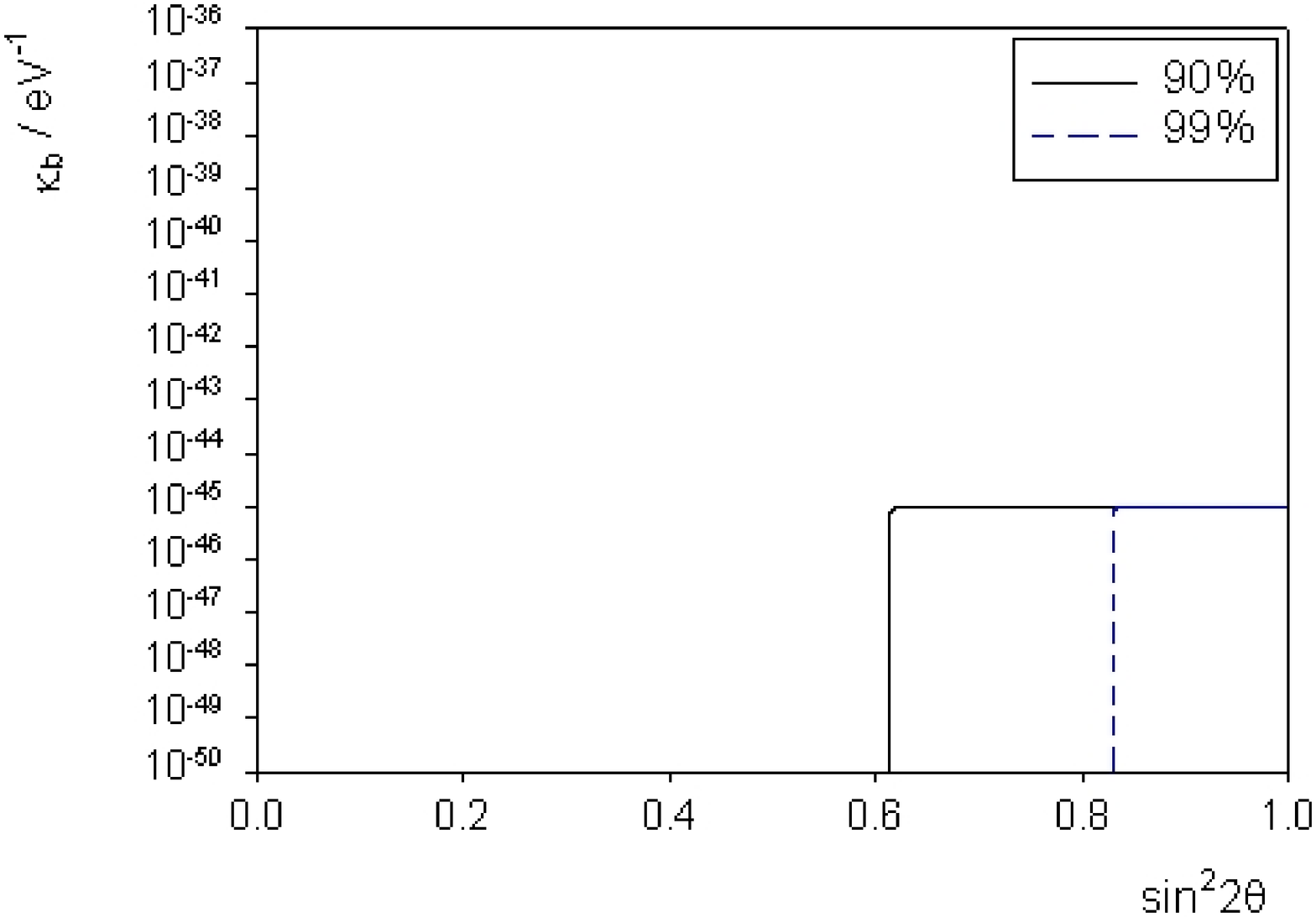} \\
\includegraphics[angle=270,width=8cm]{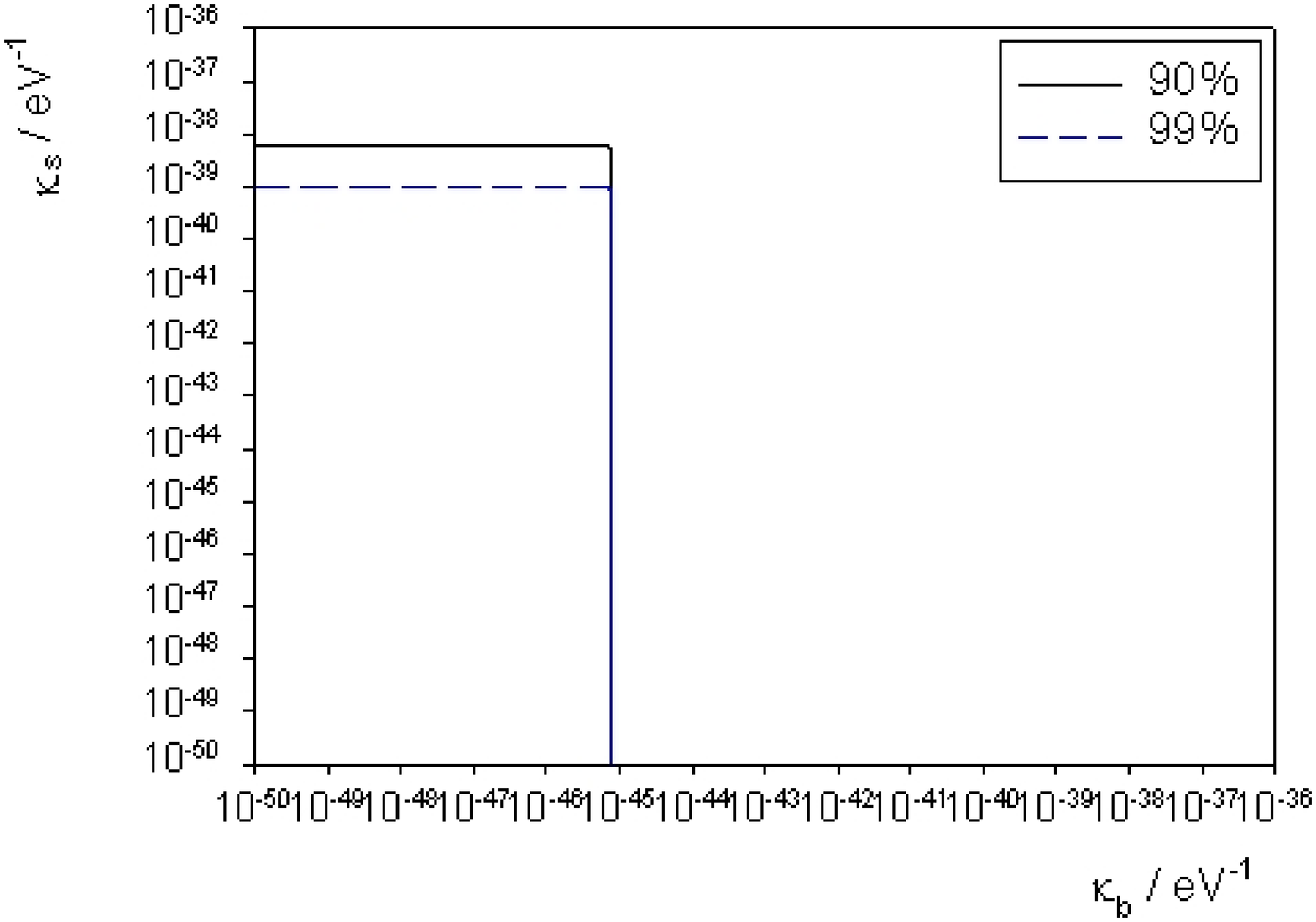}
\end{center}
\caption{Sensitivity contours at 90 and 99 percent confidence
levels, with two non-zero decoherence parameters which are
 proportional to the neutrino energy squared and given in terms
of the constants $\kappa _{s}=\kappa _{\alpha }=\kappa _{a}$ and $\kappa _{b}$.}
\label{fig:mu2smu2b}
\end{figure}

As an illustration of our results in these models, we plot in figure
\ref{fig:mu2smu2b} the sensitivity curves
for the model in which $a=\alpha \neq 0$, $b\neq 0$, $\Delta m^{2}$
is fixed at its best fit value (\ref{eq:bestfit}) \cite{fogli},
and with the quantum decoherence parameters proportional to the
neutrino energy squared, as it is in this case that we have the most stringent
bounds on quantum decoherence effects.
We find upper bounds on the sensitivity regions of
$\kappa _{s}=\kappa _{a}=\kappa _{\alpha }<10^{-38}$ eV$^{-1}$ and
$\kappa _{b}<10^{-45}$ eV$^{-1}$.
As with model 1, these bounds are very strong, and it is interesting to note
that our bound on $\kappa _{a}=\kappa _{\alpha }$ is a couple of orders of
magnitude smaller in this model than it was in model 1.

If the quantum decoherence parameters are independent of the neutrino energy,
we find upper bounds to our sensitivity regions of
$\gamma _{a}=\gamma _{\alpha }< 10^{-13}$ eV
and $\gamma _{b}< 10^{-16}$ eV, this latter value again being due to the cut-off in $b$.
For the third possibility, when the quantum decoherence parameters are
inversely proportional to the neutrino energy, our upper bounds are somewhat weaker,
at $\mu ^{2}_{\alpha }=\mu ^{2}_{a} < 10^{0}$ eV${}^{2}$, and
$\mu ^{2}_{b} < 10^{-2}$ eV${}^{2}$.

\subsubsection{Model 5}
\label{sec:modela}

\begin{figure}
\begin{center}
\includegraphics[angle=270,width=8cm]{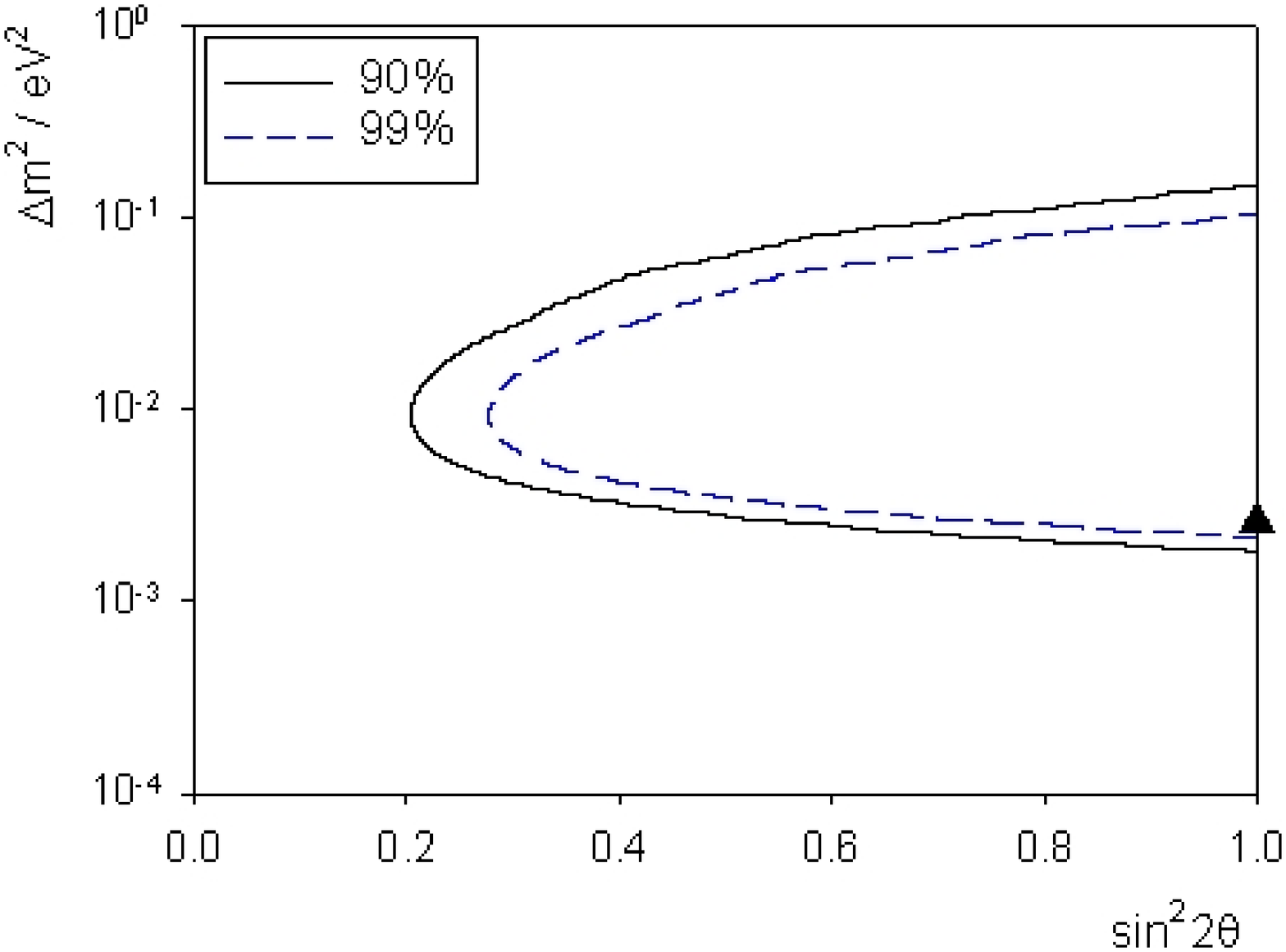}\\
\includegraphics[angle=270,width=8cm]{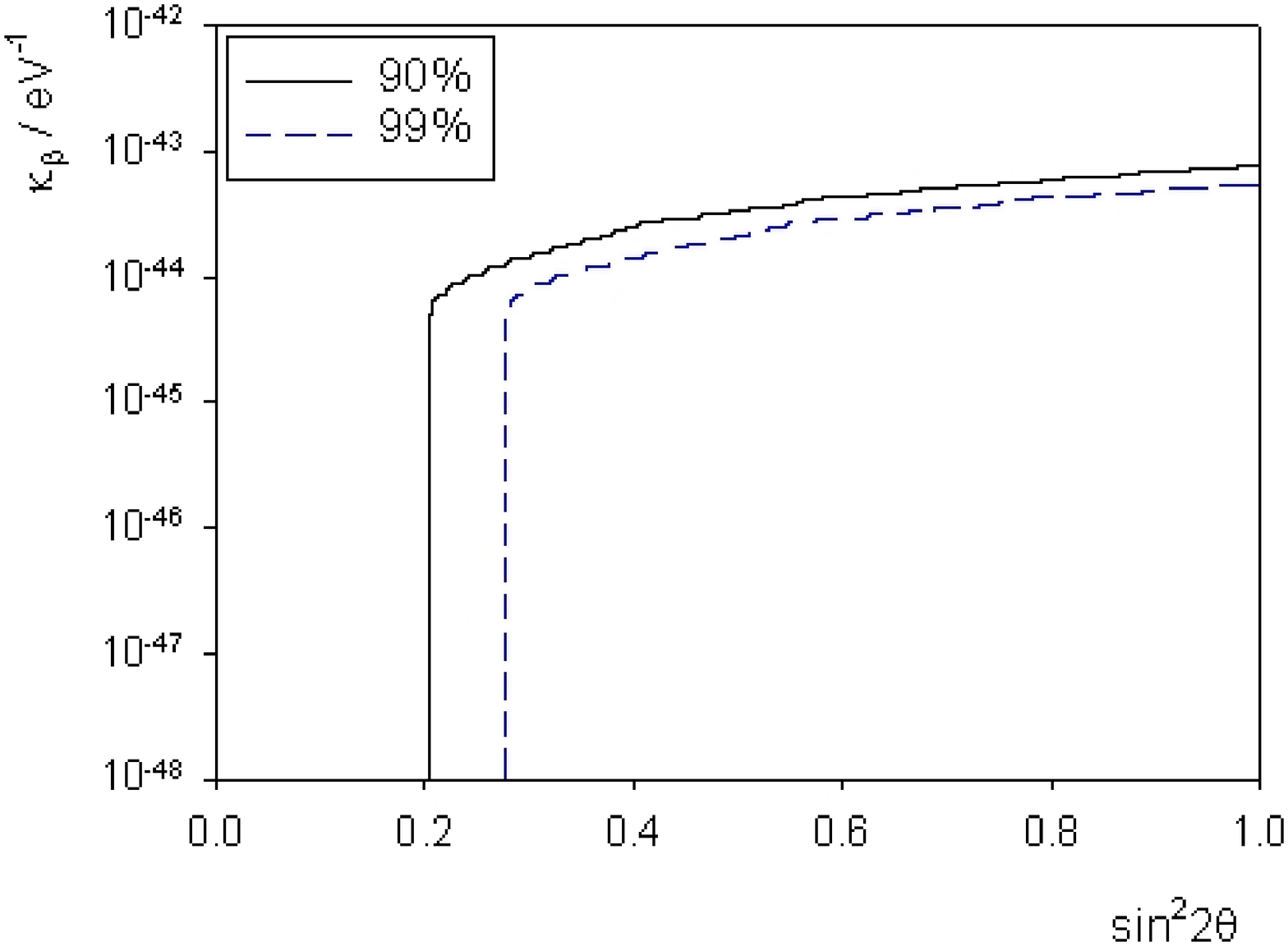} \\
\includegraphics[angle=270,width=8cm]{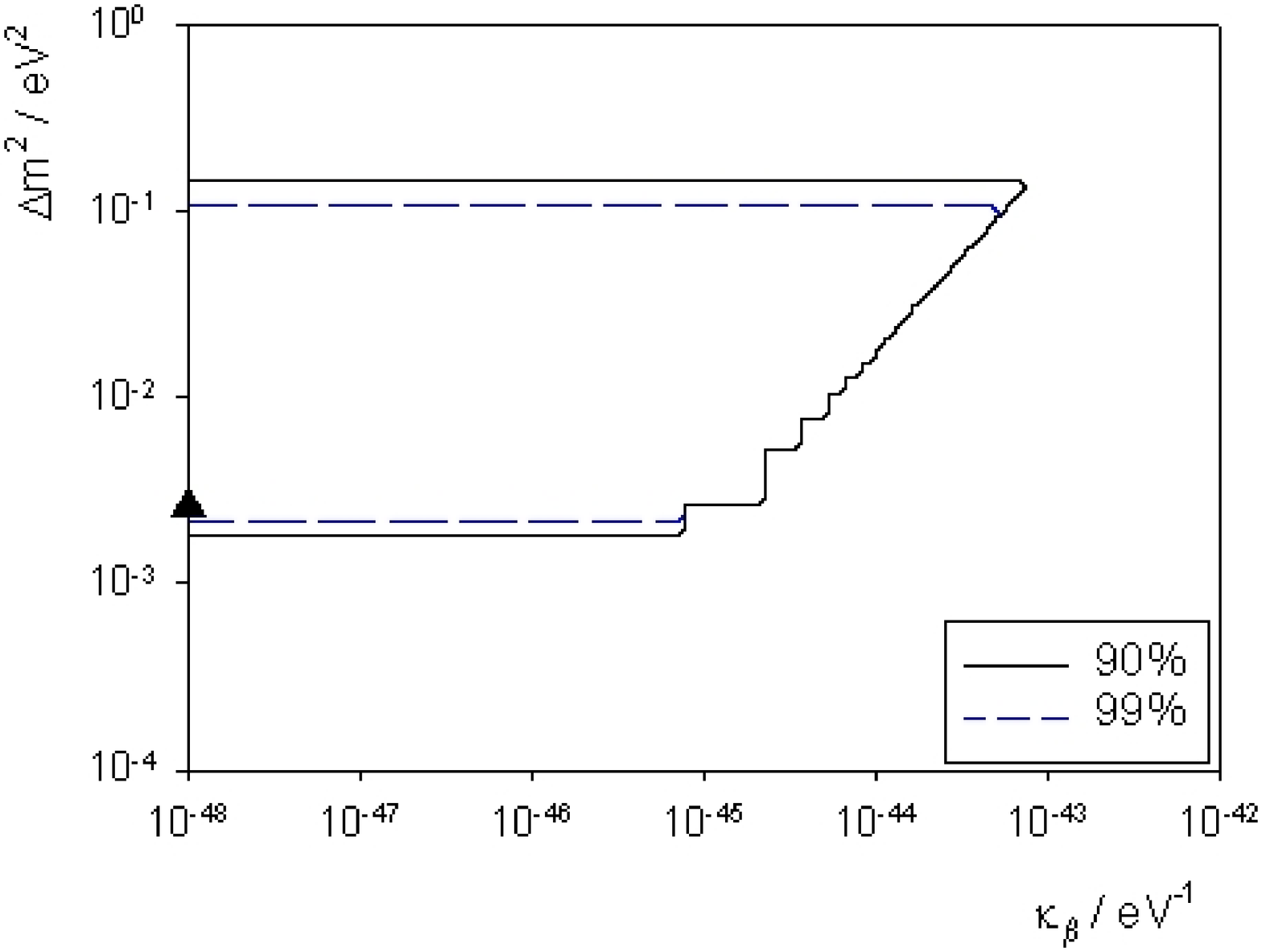}
\end{center}
\caption{Sensitivity contours at 90 and 99 percent confidence
levels, for model 5 with both standard oscillations and non-zero decoherence parameter
$\kappa_{\beta}$.  The triangle shows the experimental point of best fit
for $\Delta m^{2}$ \cite{fogli}.}
\label{fig:gambeta}
\end{figure}

To illustrate the typical sorts of plots we get for the sensitivity curves, we show in figure
\ref{fig:gambeta} the sensitivity contours for this model in the case in which the quantum decoherence
parameter proportional to the neutrino energy squared and given by $\kappa _{\beta }$.
There is again a cut-off in the
parameter $\kappa _{\beta }$ due to the probability
(\ref{eqn:probbeta}) becoming imaginary, and this can be clearly
seen in the plots.
Our upper bound on $\kappa _{\beta }$ is $10^{-43}$ eV${}^{-1}$ in this case.
The plots shown in figure
\ref{fig:gambeta} are typical of those found in many of our models
for the different quantum decoherence parameters, when we also
vary $\Delta m^{2}$.

\begin{figure}
\begin{center}
\includegraphics[angle=270,width=8cm]{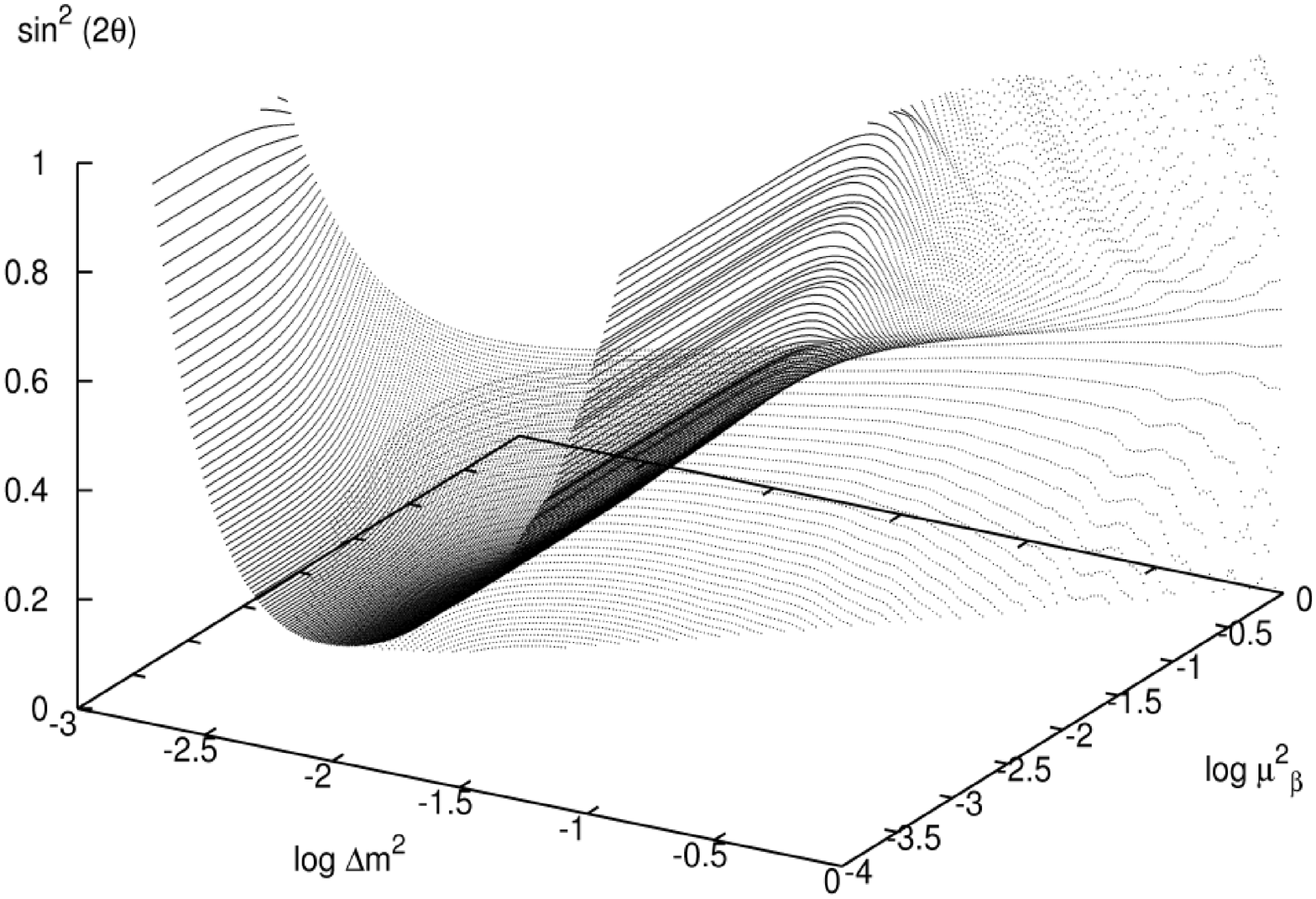}
\end{center}
\caption{The boundary of the volume in parameter space at 90\% confidence level, for model 5
with both standard oscillations and a non-zero decoherence parameter $\mu _{\beta }^{2}$.
The sensitivity region lies above this surface.}
\label{fig:mubeta3d}
\end{figure}

When the quantum decoherence parameters are independent of the neutrino energy,
the sensitivity regions are qualitatively the same as those in figure \ref{fig:gambeta},
and we find an upper bound $\gamma _{\beta } < 10^{-15}$ eV.
However, when the quantum decoherence parameters are inversely proportional to
the neutrino energy, the situation is rather more complicated, and we need
to examine the surface in three-dimensional parameter space which bounds
the sensitivity region.
This is shown in figure \ref{fig:mubeta3d} for 90\% confidence level.
The region of interest lies above the surface in figure
\ref{fig:mubeta3d}, including the value $\sin ^{2} (2\theta )=1$.
For small values of $\mu _{\beta }^{2}$, it is clear that this
region is similar in shape to the previous figure
\ref{fig:gambeta}.
However,
for larger values of $\mu ^{2}_{\beta }$, the surface opens out
and all values of $\Delta m^{2}$ and $\sin ^{2} (2\theta )$ are
included.
Because of this, we are unable to place an upper bound
on $\mu ^{2}_{\beta }$ in this case.
This is not entirely
unexpected because ANTARES measures high energy
neutrinos, so that its ability to place bounds on effects
inversely proportional to neutrino energy is limited.

\subsubsection{Model 6}
\label{sec:modelb}

This model is somewhat different to any considered before as the
parameter $d$ is non-zero and so the oscillation probability
(\ref{eqn:probc}) has contributions from the
$\sin4\theta$ term.
However, the sensitivity regions are broadly similar to those in the
previous model.
We find the following upper bounds on the sensitivity regions:
\begin{displaymath}
\gamma _{d}<10 ^{-15} \,\,{\mbox {eV}}; \qquad \qquad \mu ^{2}_{d}
< 10^{-1} \,\,{\mbox {eV${}^{2}$}}; \qquad \qquad \kappa _{d} <
10^{-43} \,\,{\mbox {eV${}^{-1}$}} .
\end{displaymath}

\section{Comparing experimental bounds on decoherence parameters}
\label{sec:exp}

In this section we compare our upper bounds on quantum decoherence
parameters, from simulations, with existing experimental data.
Firstly, for ease of comparison, we restate, in table
\ref{table:gammukap}, the upper bounds of the
constants of proportionality $\gamma $, $\mu ^{2}$ and $\kappa $ for each of the quantum
decoherence parameters, changing the units to GeV for ease of comparison with other references.
\begin{table}
\center{\begin{tabular}{|c|c|c|c|c|c|} \hline
model  & $a$  & $|b|$ & $|d|$ & $\alpha $ & $|\beta |$\\
\hline
$\gamma({\mbox {GeV}})$
& $9\times 10^{-23}$ & $7\times 10^{-25}$ & $7\times 10^{-25}$ & $9\times 10^{-23}$ & $7\times 10^{-25}$\\
\hline $\mu^{2}({\mbox {GeV${}^{2}$}})$
& $2\times 10^{-19}$ & $6\times 10^{-19}$ & $6\times 10^{-19}$ & $2\times 10^{-19}$ & -\\
\hline $\kappa({\mbox {GeV${}^{-1}$}})$
& $4\times 10^{-26}$ & $7\times 10^{-35}$ & $7\times 10^{-35}$ & $4\times 10^{-26}$ & $7\times 10^{-35}$\\
\hline
\end{tabular}}
\caption{Table showing the upper bounds of the
constants of proportionality in the quantum decoherence
parameters from simulations for different dependences on the neutrino energy.}
\label{table:gammukap}
\end{table}
Upper bounds have been placed on the decoherence parameters
\cite{lisi,fogli} by examining data from the Super-Kamiokande and
K2K experiments.
We note than the bounds for the
$\gamma$ model ($\gamma_{\alpha} < 3.5\times 10^{-23}$ GeV)
are entirely consistent with those found in our
simulations, the bounds for the $\mu^{2}$ model
($\mu_{\alpha}^{2} < 2.44 \times 10^{-21}$ GeV$^{2}$) are
slightly better than the ANTARES bounds whereas the bounds found
for the $\kappa$ model
($\kappa_{\alpha} < 9\times 10^{-10}$ GeV$^{-1}$) are much worse than our bounds.
The ANTARES
bounds are found with high energy neutrinos whereas
Super-Kamiokande detects lower energy neutrinos and so we see that
experiments which are more sensitive to lower energy neutrinos are
better probes of the $\mu^{2}$ model whilst those with sensitivity
to higher energy neutrinos are much better at placing bounds on
the quantum gravity parameters using the $\kappa$ model.
For each quantum decoherence parameter, our upper bounds on the constant $\kappa $
are particularly strong, and, given that $\kappa $ contains one inverse power of the
Planck mass ($10 ^{19}$ GeV), are close to ruling out such effects.

The neutrino system is not the only quantum system which may be
affected by interactions with space-time foam.
Experiments
such as CPLEAR \cite{CPLEAR} have examined this problem with
neutral kaons, and found the upper bounds on the quantum
decoherence parameters to be \cite{Ellisbound}
\begin{displaymath}
a<4\times 10^{-17} \,\,{\mbox {GeV}}, \qquad
|b|<2.3\times 10^{-19} \,\,{\mbox {GeV}}, \qquad
\alpha<3.7\times 10^{-21} \,\,{\mbox {GeV}}.
\end{displaymath}
A second analysis of this
experiment took place in reference \cite{benattibound} with  six
quantum decoherence parameters.
The authors of reference
\cite{benattibound} were able to put bounds all the parameters
and they found upper bounds on the quantum decoherence parameters
to be of order $10^{-17}-10^{-18}$ GeV.
It is difficult to directly compare our results with these bounds due to the
dependence on the neutrino energy in some of our models.
However, when the quantum decoherence parameters are constants, our upper bounds
on $\gamma $ from table \ref{table:gammukap} lead to upper bounds on the decoherence
parameters themselves of
\begin{displaymath}
a=\alpha < 5\times 10^{-23}\,\,{\mbox {GeV}},
\qquad |b|=|d| = |\beta | <3\times10^{-25}\,\,{\mbox {GeV}},
\end{displaymath}
and so, it seems that the
ANTARES experiment will be able to improve the upper bounds of
these parameters with respect to the CPLEAR experiment.

Quantum decoherence has also been explored using neutron
interferometry experiments, as suggested in \cite{EHNS}.
In reference \cite{neutron}, upper bounds on
the parameters $a$ and $\alpha$ were found using data from
such an experiment:
\begin{displaymath}
a\leq1\times 10^{-22}\,\, {\mbox {GeV}} ,
\qquad \alpha\leq7.4\times10^{-22}\,\,{\mbox {GeV}}.
\end{displaymath}
Once again, our upper bounds are tighter than these values.
It is interesting to note
that in both the kaon and neutron experiments, the quantum
decoherence parameter $a \neq \alpha$ whereas in the neutrino
system, we find $a=\alpha$.
This variation arises from the
different mixing in these systems.
The other difference between the meson and neutrino experiments
is the possibility of different decoherence effects in the anti-neutrino
sector as compared with the neutrino sector \cite{barenboim1,barenboim2},
although we have not considered this possibility further in this paper.

\section{Other effects on atmospheric neutrino oscillations}
\label{sec:other}

Quantum decoherence is not the only process which may modify the
standard oscillation probability (\ref{eq:standard}) for
atmospheric neutrinos.
In this section we briefly discuss some of
the other possible phenomena, indicating how they change the
oscillation probability.
We will then compare these effects with
those coming from quantum decoherence.

The derivations of the oscillation probabilities have all been
done within the framework of standard quantum mechanics.
It has
been suggested, however, that approaching this problem from a
quantum field theory (QFT) standpoint could alter the
phenomenology \cite{blasone,hannabuss}.
This approach yields statistical averages, but if we interpret
these as probabilities, then there will be a modification of
the standard oscillation probability.
However, the magnitude of the correction is several orders
of magnitude smaller than the size of the corrections due
to quantum decoherence that we are probing here.

There are other new physics effects which may alter the
standard atmospheric neutrino oscillation probability, not
just quantum decoherence.
For example, it has been suggested that our universe may have more than the
$3+1$ dimensions we observe and that some of these extra
dimensions may be large (see \cite{Gabadadze} for an
introduction), which implies that we live on a four dimensional
hyper-surface, a brane, embedded in a higher dimensional bulk.
Adding large extra dimensions gives rise to the possibility that
some particles may propagate off our brane through the bulk and it
has been suggested that one of these particles may be a sterile
neutrino into which the three familiar neutrinos could oscillate.
If we consider the mixing between the three standard neutrinos and
the sterile neutrino to simply be an extension of the three
neutrino system to a four neutrino system,
and assuming that $E_{1,2} << E_{3} << E_{4}$, then the
probability a muon neutrino oscillates into a tau neutrino is
\begin{eqnarray*}
\nonumber P[\nu_{\mu} \rightarrow \nu_{\tau}]
&=& 4U_{\mu 3}U_{\tau 3}(U_{\mu 3}U_{\tau 3}+U_{\mu 4}U_{\tau
4})\sin^{2}\left[\frac{\Delta m^{2}_{32}}{4E}L\right]\\
& & +4U_{\mu 4}^{2}U_{\tau 4}^{2}\sin^{2}\left[\frac{\Delta
m^{2}_{43}}{4E}L\right].
\end{eqnarray*}
Here, the probability has an extra term as we now have two large
mass differences.
However, the quantities $U_{xi}$ are constants, independent of $E$ and $L$ and
hence, this simple model gives rise to different types of terms
than those found in equation (\ref{eq:genprob}).
Similar corrections to the two-neutrino oscillation probability also arise
if we consider normal three-neutrino oscillations, but in this case the
corrections are very small.

It has, however, also been suggested that the sterile neutrino may not mix in the
standard way  and that the
the oscillation probability is instead \cite{Barbieri}:
\begin{eqnarray}
\label{eqn:LEDprob}
\nonumber
P[\nu_{a} \rightarrow \nu_{b}] &=&
4U_{a3}^{2}U_{b3}^{2}e^{-\frac{\pi^{2}\xi_{3}^{2}}{2}}\sin^{2}
\left[\frac{\Delta m_{32}^{2}}{4E}L\right]\\
\nonumber & &
+2\sum_{l=1}^{3}U_{a3}U_{b3}U_{al}U_{bl}\sum_{n=1}^{\infty}
U_{0n}^{2}\cos\left[\frac{(\lambda_{n}^{2}-\xi_{ll}^{2})L}{2ER^{2}}\right]\\
& &+U_{a3}^{2}U_{b3}^{2}\left|
\sum_{n=1}^{\infty}U_{0n}^{2}e^\frac{-i\lambda_{n}^{2}}{2ER^{2}}L\right|^{2}
\end{eqnarray}
where $R$ is the size of the extra dimensions, $\xi _{i}=m_{i}R$ are assumed to be small,
$ll=2$ if $l=1,2$ or $ll=3$ if $l=3$ and the $U$'s are constants.
This oscillation probability has a different form to that in
equation (\ref{eq:genprob}) as the exponential term multiplies the
$\sin^{2}\left[\frac{\Delta m^{2}}{4E}L\right]$ as opposed to the
$\cos$ term. Also, if we assume that $R$ does not change with
time, then the exponential term in equation (\ref{eqn:LEDprob}) is
a constant whereas the exponential term in equation
(\ref{eq:genprob}) is a function of $E$ and $L$.

Another way in which quantum gravity is expected to modify neutrino physics is
through violations of Lorentz invariance \cite{nick,hugo}.
As with non-standard sterile neutrinos, this leads to modifications
of the oscillation probability which have a different form from those
due to quantum decoherence.
We will examine in a separate publication the precise nature of these effects
on atmospheric neutrino oscillations and whether these can be probed by ANTARES.
Other types of Lorentz-invariance and CPT-violating effects (also known as ``Standard Model Extensions'' (SMEs))
have been studied by Kostelecky and Mewes \cite{kostelecky}, and again we shall return to these
for ANTARES in the future.
One side-effect of SMEs is that, even for atmospheric neutrinos, matter interactions
can become significant.
However, it has been shown that these can safely be ignored for neutrino energies above the scale of
tens of GeV \cite{matter}, which is the energy range of interest to us here.

Finally, more conventional physics may have an effect on atmospheric neutrino oscillations as
measured by a neutrino telescope.
For example, uncertainties in the energy and path length of the neutrinos may lead to modifications
of the oscillation probability which are similar in form to those arising from quantum decoherence
\cite{ohlsson,barenboim1}
and interactions in the Earth may also be a factor.
However, both these effects decrease as the neutrino energy increases, whereas our strongest results are for
quantum decoherence effects which increase with neutrino energy.

\section{Conclusions}
\label{sec:conc}

In the absence of a complete theory of quantum gravity, phenomenological signatures of new
physics are useful for experimental searches.
In this article we have examined one expected effect of quantum gravity: quantum decoherence.
We have considered a general model of decoherence applied to atmospheric neutrinos, and simulated
the sensitivity of the ANTARES neutrino telescope to the decoherence parameters in our model.
We find that high-energy atmospheric neutrinos are able to place very strict constraints on
quantum decoherence when the decoherence parameters are proportional to the neutrino energy squared,
even though such effects are suppressed by at least one power of the Planck mass.
It should be stressed that although our results here are for the ANTARES experiment, we expect that
other high-energy neutrino experiments (such as ICECUBE and AMANDA) should also be able to probe
these models.

\section*{Acknowledgements}
This work was supported by PPARC, and
DM is supported by a PhD studentship from the University of Sheffield.
We would like to thank Keith Hannabuss and Nick Mavromatos for helpful discussions.


\begin{thebibliography}{10}

\bibitem{smolin}
L. Smolin, {\em {How far are we from the quantum theory of
gravity?}}, Preprint {\tt {hep-th/0303185}}.

\bibitem{amelino1}
G. Amelino-Camelia, {\em {Introduction to quantum gravity phenomenology}},
Preprint {\tt {gr-qc/0412136}}.

\bibitem{rs}
N. Arkani-Hamed, S. Dimopoulos and G.R. Dvali,
{\em {Phys. Lett.}} {\bf {B429}} 263 (1998);
\newline
L. Randall and R. Sundrum,
{\em {Phys. Rev. Lett.}} {\bf {83}} 3370 (1999);
\newline
L. Randall and R. Sundrum,
{\em {Phys. Rev. Lett.}} {\bf {83}} 4690 (1999).

\bibitem{EHNS}
J.R. Ellis, J.S. Hagelin, D.V. Nanopoulos and M. Srednicki,
{\em {Nucl. Phys.}} {\bf {B241}} 381 (1984).

\bibitem{benatti}
F. Benatti and R. Floreanini,
{\em {J. High. Energy Phys.}} {\bf {02}} 032 (2000).

\bibitem{liu}
Y. Liu, L-Z. Hu and M-L. Ge,
{\em {Phys. Rev.}} {\bf {D56}} 6648 (1997).

\bibitem{ma}
F-C. Ma and H-M. Hu,
{\em {Testing quantum mechanics in neutrino oscillation}},
Preprint {\tt {hep-ph/9805391}} (1998).

\bibitem{chang}
C-H. Chang, W-S. Dai, X-Q. Li, Y. Liu, F-C. Ma and Z-J. Tao,
{\em {Phys. Rev.}} {\bf {D60}} 033006 (1999).

\bibitem{klapdor}
H.V. Klapdor-Kleingrothaus, H. Pas and U. Sarkar,
{\em {Eur. Phys. J.}} {\bf {A8}} 577 (2000).

\bibitem{lisi}
E. Lisi, A. Marrone and D. Montanino,
{\em {Phys. Rev. Lett.}} {\bf {85}} 1166 (2000).

\bibitem{fogli}
G.L. Fogli, E. Lisi, A. Marrone and D. Montanino,
{\em {Phys. Rev.}} {\bf {D67}} 093006 (2003).

\bibitem{SKspectra}
Y. Ashie et. al. (Super-Kamiokande Collaboration)
{\em {Phys. Rev. Lett.}} {\bf {93}} 101801 (2004).

\bibitem{ANTARES}
E.V. Korolkova (for the ANTARES collaboration),
{\em {Nucl. Phys. Proc. Suppl.}} {\bf {136}} 69 (2004).

\bibitem{nick}
N.E. Mavromatos, {\em {On CPT symmetry: Cosmological, quantum-gravitational
and other possible violations and their phenomenology}},
Preprint {\tt {hep-ph/0309221}}.
\newline
N.E. Mavromatos,
{\em {Neutrinos and the phenomenology of CPT violation}},
Preprint {\tt {hep-ph/0402005}}.
\newline
N.E. Mavromatos, {\em {CPT violation and decoherence in quantum gravity}},
Preprint {\tt {gr-qc/0407005}}.

\bibitem{brustein}
R. Brustein, D. Eichler and S. Foffa,
{\em {Phys. Rev.}} {\bf {D65}} 105006 (2002).

\bibitem{hugo}
J. Alfaro, H.A. Morales-T\'ecotl and L.F. Urrutia,
{\em {Phys. Rev. Lett.}} {\bf {84}} 2318 (2000).

\bibitem{king}
S. Choubey and S.F. King,
{\em {Phys. Rev.}} {\bf {D67}} 073005 (2003).

\bibitem{ELMN}
J.R. Ellis, J.L. Lopez, N.E. Mavromatos and D.V. Nanopoulos, {\em
{Phys. Rev.}} {\bf {D53}} 3846 (1996).

\bibitem{benattibound}
F. Benatti and R. Floreanini, {\em{Phys. Lett.}} {\bf{B401}} 337
(1997).

\bibitem{AMANDA}
E. Andres et. al. {\em{Nature}} {\bf{410}} 441 (2001).

\bibitem{ICECUBE}
J. Ahrens et. al., {\em{Astropart. Phys.}} {\bf{20}} 507 (2004).

\bibitem{NESTOR}
S.E. Tzamarias, {\em{Nucl. Instrum. Meth.}} {\bf{A502}} 150
(2003).

\bibitem{KAMLANDspectra}
T. Araki et. al. (KAMLAND collaboration)
{\em {Phys. Rev. Lett.}} {\bf {94}} 081801 (2005).

\bibitem{MINOS}
P.J. Litchfield, {\em {Nucl. Instrum. Meth.}} {\bf {A451}} 187 (2000).

\bibitem{OPERA}
M. Komatsu, {\em {Nucl. Instrum. Meth.}} {\bf {A503}} 124 (2003).

\bibitem{blennow}
M. Blennow, T. Ohlsson and W. Winter,
{\em {J. High Energy Physics}} {\bf {0506}} 049 (2005).

\bibitem{wheeler}
J.A. Wheeler,
{\em {Superspace and the nature of quantum geometrodynamics}},
in {\em {Battelle Rencontres: 1967 Lectures in Mathemetical Physics}}
edited by C. DeWitt and J.A. Wheeler (W. A. Benjamin, New York, 1968).

\bibitem{hawking1}
S.W. Hawking, {\em {Commun. Math. Phys.}} {\bf {43}} 199 (1975).

\bibitem{hawking2}
S.W. Hawking, {\em {Commun. Math. Phys.}} {\bf {87}} 395 (1982).

\bibitem{lindblad}
G. Lindblad, {\em {Commun. Math. Phys.}} {\bf {48}} 119 (1976).

\bibitem{GRW}
G.C. Ghirardi, A. Rimini and T. Weber, {\em {Phys. Rev.}} {\bf {D34}} 470 (1986).

\bibitem{decobook}
E. Joos, H.D. Zeh, C. Kiefer, D. Giulini, J. Kupsch and
I.-O. Stamatescu,
{\em {Decoherence and the appearance of a classical world in quantum theory}},
Springer (1996).

\bibitem{gambini}
R. Gambini, R.A. Porto and J. Pullin, {\em {Class. Quantum Gravity}} {\bf {21}} L51 (2004).

\bibitem{ohlsson}
T. Ohlsson, {\em {Phys. Lett.}} {\bf {B502}} 159 (2001).

\bibitem{barenboim1}
G. Barenboim and N.E. Mavromatos, {\em {Phys. Rev.}} {\bf {D70}} 093015 (2004).

\bibitem{barenboim2}
G. Barenboim and N.E. Mavromatos,
{\em {J. High Energy Physics}} {\bf {0501}}  034 (2005).

\bibitem{winstanley}
J.R. Ellis, N.E. Mavromatos, D.V. Nanopoulos and E. Winstanley,
{\em {Mod. Phys. Lett.}} {\bf {A12}} 243 (1997).
\newline
J.R. Ellis, N.E. Mavromatos and D.V. Nanopoulos,
{\em {Mod. Phys. Lett.}} {\bf {A12}} 1759 (1997).
\newline
J.R. Ellis, P. Kanti, N.E. Mavromatos, D.V. Nanopoulos and E. Winstanley,
{\em {Mod. Phys. Lett.}} {\bf {A13}} 303 (1998).

\bibitem{myers}
R.C. Myers and M. Pospelov, {\em {Phys. Rev. Lett.}} {\bf {90}} 211601 (2003).

\bibitem{adler}
S.L. Adler, {\em {Phys. Rev.}} {\bf {D62}} 117901 (2000).

\bibitem{atmflux}
G. Battistoni, A. Ferrari, T. Montaruli and P.R. Sala,
{\em {Astropart. Phys.}} {\bf {19}} 269 (2003); {\em {ibid}} {\bf {19}} 291 (2003).
\newline
C.G.S. Costa, {\em {Astropart. Phys.}} {\bf {16}} 193 (2001).

\bibitem{dan}
D. Hooper, D. Morgan and E. Winstanley,
{\em {Phys. Lett.}} {\bf {B609}} 206 (2005).
\newline
D. Hooper, D. Morgan and E. Winstanley,
{\em {Phys. Rev.}} {\bf {D72}} 065009 (2005).

\bibitem{K2K}
M.H. Ahn et. al. (K2K Collaboration), {\em{Phys. Rev. Lett.}}
{\bf{90}} 041801 (2003).

\bibitem{KamLAND}
K. Eguchi et. al. (KamLAND Collaboration), {\em{Phys. Rev. Lett.}}
{\bf{90}} 021802 (2003).

\bibitem{CHOOZ}
M. Apollonio et. al., {\em {Phys. Lett.}} {\bf{B420}} 397 (1998).

\bibitem{antosc}
F. Blondeau and L. Moscoso (ANTARES collaboration),
{\em {Detection of atmospheric neutrino oscillations with a 0.1 km${}^{2}$ detector : the case for ANTARES}}
NOW 98 conference, Amsterdam, Netherlands (1998).
\newline
J. Brunner (ANTARES collaboration),
{\em {Measurement of neutrino oscillations with neutrino telescopes}},
15th International Conference On Particle And Nuclei (PANIC 99), Uppsala, Sweden (1999).
\newline
C. Carloganu (ANTARES collaboration),
{\em {Nucl. Phys. Proc. Suppl.}} {\bf {100}} 145 (2001).

\bibitem{bartol}
T.K. Gaisser, M. Honda, P. Lipari, T. Stanev,
{\em {Primary spectrum to 1 TeV and beyond}},
International Conference on Cosmic Rays (ICRC 2001), Hamburg, Germany (2001).

\bibitem{K2Kspectra}
E. Aliu et. al. (K2K Collaboration),
{\em {Phys. Rev. Lett.}} {\bf {94}} 081802 (2005).

\bibitem{SK}
Y. Fukuda et. al. (SuperKamiokande Collaboration), {\em{Phys.
Lett.}} {\bf{B433}} 9 (1998).

\bibitem{CPLEAR}
CPLEAR Collaboration, {\em{Phys. Rep.}} {\bf{374}} 165 (2003).

\bibitem{Ellisbound}
J. Ellis and CPLEAR Collaboration, {\em{Phys. Lett.}} {\bf{B364}}
239 (1995).

\bibitem{neutron}
F. Benatti and R. Floreanini, {\em{Phys. Lett.}} {\bf{B451}} 422
(1999).

\bibitem{blasone}
M. Blasone, P.A. Henning and G. Vitiello,
{\em {Phys. Lett.}} {\bf {B451}} 140 (1999).
\newline
M. Blasone and G. Vitiello, {\em{Quantum field theory of particle
mixing and oscillations}}, Preprint {\tt {hep-ph/0309202}}.

\bibitem{hannabuss}
K.C. Hannabuss and D.C. Latimer,
{\em {J. Phys.}} {\bf {A33}} 1369 (2000).
\newline
K.C. Hannabuss and D.C. Latimer,
{\em {J. Phys.}} {\bf {A36}} L69 (2003).

\bibitem{Gabadadze}
G. Gabadadze, {\em{ICTP Lectures on Large Extra Dimensions}},
Preprint {\tt{hep-ph/0308112}}.

\bibitem{Barbieri}
R. Barbieri, P. Creminelli and A. Strumia, {\em {Nucl. Phys.}}
{\bf{B585}} 28 (2000).

\bibitem{kostelecky}
V.A. Kostelecky and M. Mewes,
{\em {Phys. Rev.}}  {\bf {D69}} 016005 (2004).
\newline
V.A. Kostelecky and M. Mewes,
{\em {Phys. Rev.}} {\bf {D70}} 031902 (2004).

\bibitem{matter}
M. Jacobson and T. Ohlsson,
{\em {Phys. Rev.}} {\bf {D69}} 013003 (2004).

\end{thebibliography}
\end{document}